\documentclass[reprint, twocolumn, amsfonts, amsmath, amssymb, superscriptaddress, nofootinbib, prx]{revtex4-2}

\usepackage{siunitx}
\usepackage[inline]{enumitem}
\usepackage{comment}
\usepackage{graphicx}
\usepackage[normalem]{ulem}
\usepackage{mathtools}
\usepackage[dvipsnames]{xcolor}
\usepackage[linkcolor=Blue,citecolor=Blue,urlcolor=Blue,colorlinks=true]{hyperref}

\usepackage[version=4]{mhchem}
\mhchemoptions{mathfontname=mathsf}

\newif\ifshownotes
\ifshownotes
	\newcommand{\note}[3]{{\color{#2}[#1: #3]}}
    
    \newcommand{\del}[3]{\textbf{\color{#2}\sout{#3}}}
    \newcommand{\eqdel}[3]{\textbf{\color{#2}#3}}
\else
	\newcommand{\note}[3]{}
	\newcommand{\del}[3]{}
	\newcommand{\eqdel}[3]{}
    
\fi

\renewcommand{\vec}{\boldsymbol}

\begin{document}

\title{Self-consistent sharp interface theory of active condensate dynamics}
\author{Andriy Goychuk}
\email{andriy@goychuk.me}
\affiliation{A.G. and L.D. contributed equally.}
\affiliation{Arnold Sommerfeld Center for Theoretical Physics and Center for NanoScience, Department of Physics, Ludwig-Maximilians-Universit\"at M\"unchen, Theresienstra\ss e 37, D-80333 M\"unchen, Germany}
\affiliation{Institute for Medical Engineering and Science, Massachusetts Institute of Technology, Cambridge, MA 02139, United States}
\author{Leonardo Demarchi}
\affiliation{A.G. and L.D. contributed equally.}
\affiliation{Arnold Sommerfeld Center for Theoretical Physics and Center for NanoScience, Department of Physics, Ludwig-Maximilians-Universit\"at M\"unchen, Theresienstra\ss e 37, D-80333 M\"unchen, Germany}
\affiliation{Sorbonne Université, CNRS, Institut de Biologie Paris-Seine (IBPS), Laboratoire Jean Perrin (LJP), F-75005 Paris, France}
\author{Ivan Maryshev}
\affiliation{Arnold Sommerfeld Center for Theoretical Physics and Center for NanoScience, Department of Physics, Ludwig-Maximilians-Universit\"at M\"unchen, Theresienstra\ss e 37, D-80333 M\"unchen, Germany}
\author{Erwin Frey}
\email{frey@lmu.de}
\affiliation{Arnold Sommerfeld Center for Theoretical Physics and Center for NanoScience, Department of Physics, Ludwig-Maximilians-Universit\"at M\"unchen, Theresienstra\ss e 37, D-80333 M\"unchen, Germany}
\affiliation{Max Planck School Matter to Life, Hofgartenstraße 8, D-80539 M\"unchen, Germany}

\date{\today}

\begin{abstract}
Biomolecular condensates help organize the cell cytoplasm and nucleoplasm into spatial compartments with different chemical compositions.
A key feature of such compositional patterning is the local enrichment of enzymatically active biomolecules which, after transient binding via molecular interactions, catalyze reactions among their substrates.
Thereby, biomolecular condensates provide a spatial template for non-uniform concentration profiles of substrates.
In turn, the concentration profiles of substrates, and their molecular interactions with enzymes, drive enzyme fluxes which can enable novel non-equilibrium dynamics.
To analyze this generic class of systems, with a current focus on self-propelled droplet motion, we here develop a self-consistent sharp interface theory.
In our theory, we diverge from the usual bottom-up approach, which involves calculating the dynamics of concentration profiles based on a given chemical potential gradient. 
Instead, reminiscent of control theory, we take the reverse approach by deriving the chemical potential profile and enzyme fluxes required to maintain a desired condensate form and dynamics.
The chemical potential profile and currents of enzymes come with a corresponding power dissipation rate, which allows us to derive a thermodynamic consistency criterion for the passive part of the system (here, reciprocal enzyme-enzyme interactions).
As a first use case of our theory, we study the role of reciprocal interactions, where the transport of substrates due to reactions and diffusion is, in part, compensated by redistribution due to molecular interactions.
More generally, our theory applies to mass-conserved active matter systems with moving phase boundaries.
\end{abstract}

\maketitle

Biomolecular condensates contribute to intracellular organization~\cite{Brangwynne2009, Hyman2014, Banani2017, Lyon2020, Alberti2019, Shin2017, Choi2020} by controlling the local chemical composition.
The underlying mechanism, where a liquid mixture phase separates according to differences in the interactions among its components, enables condensates to naturally buffer molecules~\cite{Klosin2020} and compartmentalize reactions~\cite{Lyon2020}.
Short-ranged interactions give rise to an interfacial tension between the different phases, which thereby gradually coarsen into a single droplet through Ostwald ripening~\cite{Ostwald1900, LifshitzSlyozov1961, Wagner1961, Bray1994}.
For systems in thermal equilibrium, this coarsening process can only be arrested through long-ranged reciprocal interactions, for example due to non-local elastic stresses in polymeric gels~\cite{Qiang2024} or electrostatic interactions in block copolymer melts~\cite{Liu1989} as well as charged droplets~\cite{Rayleigh1882, Deserno2001}.
In contrast, cells combine phase separation with a myriad of active processes which consume chemical energy~\cite{ZwickerReview2022} to fuel important cellular tasks such as gene transcription~\cite{Hnisz2017, Sabari2018, Shrinivas2019, Henninger2021}, splicing~\cite{Faber2022}, ribosomal subunit assembly~\cite{Lafontaine2021}, or midcell localization during cell division~\cite{Schumacher2017}.
The underlying irreversible chemical reactions, and resulting mass currents, enable `active droplets' to exhibit a wealth of novel dynamics not encountered in thermal equilibrium~\cite{Shin2017, Banani2017, Weber2019}.

Previous studies demonstrated that continuous turnover of condensate material via chemical reactions~\cite{Weber2019, Glotzer1994, Glotzer1995, Christensen1996, Carati1997, Zwicker2015, Lamorgese2016, Zwicker2016, Wurtz2018, Li2020, Kirschbaum2021}, and the resulting material fluxes, enable multi-droplet coexistence~\cite{Carati1997, Zwicker2015, Wurtz2018, Li2020, Kirschbaum2021} and droplet division~\cite{Zwicker2016}.
These phenomena can be explained by a formal mapping to a micro-phase separating system with long-ranged interactions~\cite{Li2020, Kumar2023}.
Here, building on our previous work~\cite{Demarchi2023}, we study a different class of systems which are constrained by mass conservation.
We consider enzymes that spontaneously phase separate, or localize to an already existing condensate, and regulate chemical reactions among other molecules.
To that end, enzymes transiently bind substrates and catalytically lower the activation barrier for converting these substrates into products.
For example, in the bacterium \emph{Myxococcus xanthus}, a mobile cluster of PomX and PomY proteins bound to the nucleoid regulates the ATP-dependent cycling of PomZ between two conformations~\cite{Schumacher2017, Bergeler2018, Kober2019, Hanauer2021}.
As a second example, consider transcriptional condensates in the eukaryotic cell nucleus, which enrich RNA Polymerase II, transcription factors, Mediator, and other proteins~\cite{Hnisz2017, Shrinivas2019}.
As RNA Polymerase II assembles RNA from individual nucleotides, attractive electrostatic interactions between the negatively charged RNA and the positively charged transcription factors favor further condensate growth.
At high RNA concentrations, however, repulsive interactions due to volume exclusion lead to condensate dissolution~\cite{Henninger2021, Schede2023, Natarajan2023}.
Such feedback mechanisms between active reactions and passive phase separation are ubiquitous in the complex intracellular environment, and in membraneless organelles such as the nucleolus~\cite{Lafontaine2021}.

For this generic class of systems, the nonequilibrium chemical reactions catalyzed by enzymes give rise to substrate concentration gradients which, in turn, drive enzyme fluxes via reciprocal interactions.
As we have previously shown, these two coupled mechanisms can lead to condensate self-propulsion, positioning, interrupted coarsening, and division~\cite{Demarchi2023}.
Here, we significantly extend our understanding of such systems by developing a theoretical framework to determine the velocity of self-propelling droplets and phase boundaries in arbitrary dimensions.
We achieve this by deriving a self-consistent sharp interface theory, which proceeds in two steps.
At the outset, we determine the enzyme currents and chemical potential profile necessary to maintain the droplet in a nonequilibrium steady state with constant shape and velocity.
These currents, in the inferred chemical potential landscape, define the rate of free energy dissipation required to induce droplet motion with a given velocity.
This energy is injected by active processes, which perform work on the system by applying an active force field.
Since the active force field accounts for all sources and sinks of energy, the chemical potential profile must correspond to reciprocal enzyme-enzyme interactions.
In this sense, the enzymes act as a purely passive phase-separating component.
Following this, we derive a thermodynamic consistency relation grounded in the fundamental principle that passive systems reach thermal equilibrium by minimizing their free energy.
We apply our theoretical framework to a model where condensate motion is driven by interactions between the various chemical species in the bulk and does not require viscous hydrodynamic coupling.
This stands in contrast to previous studies, where liquid droplets propelled through Marangoni flows in viscous fluids~\cite{Review::Maass2016, Review::Dwivedi2022, MichelinReview2022}, through active stresses on surfaces~\cite{Joanny2012, Khoromskaia2015, Whitfield2016, Kree2017, Yoshinaga2019, Trinschek2020}, or by altering their wetting properties~\cite{Thiele2004, John2005}.
While we have previously considered a scenario where enzyme-substrate and enzyme-product interactions are weak, we here relax this assumption.
To that end, we explicitly account for the effect of enzyme concentration gradients on the diffusion of substrates and products, and study how this affects the motion of phase boundaries.
Our analysis further elucidates the range of parameters in which droplet self-propulsion can occur, and identifies a subcritical bifurcation as a function of the mobilities of enzymes as well as substrates and products.

The outline of the present article is as follows.
In Sec.~\ref{sec:model_framework}, we discuss the theoretical framework of describing a mass-conserving multi-component mixture containing enzymes which phase-separate through attractive interactions among themselves, and also interact with other molecular species such as substrates and products.
Moreover, we give an account of non-equilibrium reactions that can give rise to inhomogeneous concentration profiles of substrates and products, thereby inducing an inhomogeneous driving force on the enzymes.
As the central contribution of our work, in Sec.~\ref{sec:self_consistency_relation_generalized} we show how the core of the model can be reduced to an implicit description that is independent of the specific details of the interactions and reactions that we have introduced in the preceding discussion.
In doing so, we derive self-consistency relations that characterize the enzyme currents, chemical potential profile, and the resulting droplet velocity for an arbitrary inhomogeneous driving force.
We test our theory and demonstrate that liquid-like droplets are more readily set in motion than solid-like condensates.
In Sec.~\ref{sec:reciprocal_interactions}, we discuss the impact of reciprocal interactions.
To that end, as discussed in detail in Appendix~\ref{sec:substrate_interaction_profile_reciprocal}, we first show how reciprocal interactions can give rise to discontinuities and kinks in the substrate and product concentration profiles, how the reactants are redistributed, and how the resulting concentration profiles can be determined.
With these tools in hand, we derive the conditions to observe droplet self-propulsion in the presence of weakly or strongly reciprocal interactions, thereby complementing our previous work~\cite{Demarchi2023}.
We find that the absence or presence of reciprocal interactions determines if the onset of self-propulsion is continuous, or discontinuous.

\section{Enzyme-enriched condensates} 
\label{sec:model_framework}

\subsection{Thermodynamic currents}

Biomolecular condensates can consist of many interacting components~\cite{Shin2017}.
In this study, we investigate regular mixtures of enzyme, substrate, and product molecules in solution.
We assume that the enzymes spontaneously phase separate in thermal equilibrium, and that they act as scaffolds which transiently bind substrates and products.
Hence, in the following, the condensate is synonymous to regions where the enzyme concentration $c(\mathbf{x},t)$ is high.
Phase separation is driven by a competition between enzyme insolubility and entropy, which we encode into a Ginzburg-Landau expansion of the free energy density,
\begin{equation}
\label{eq:freeenergy_cahnhilliard}
    f_0(c) 
    = 
    - 
    \frac{r}{2} \, 
    (c-\tilde c)^2 
    +
    \frac{u}{4} \, 
    (c-\tilde c)^4 
    + 
    \frac{\kappa}{2} \, |\boldsymbol\nabla c|^2 \, ,
\end{equation}
near the critical point.
The control parameter ${r>0}$ indicates attractive enzyme-enzyme interactions, which lead to phase separation, and measures the distance from the critical point.
The parameter ${u > 0}$ is required to thermodynamically stabilize the system, so that the free energy density has the form of a double-well potential.
Finally, a positive stiffness parameter, ${\kappa > 0}$, penalizes concentration gradients.

We are interested in a scenario where the concentrations of substrates, $s(\mathbf{x},t)$, and products, $p(\mathbf{x},t)$, or the molecular volumes $\nu$ of these particles, are small.
In that case, substrates and products are associated with the free energy density of an ideal mixture, ${f_\text{I} (\varrho) = k_\text{B} T \, \varrho \log(\varrho \, \nu)}$, where ${\varrho \in \{ s,p \}}$, and  $k_\text{B} T$ is the thermal energy.
Finally, we parameterize enzyme-substrate and enzyme-product interactions with Flory-Huggins (FH) parameters $\chi_s$ and $\chi_p$, respectively. 
Summing up all of these contributions, the thermodynamics of the mixture are characterized by the effective free energy functional,
\begin{equation} 
    \mathcal{F} 
    = 
    \int\! d^dx \, 
    \big[ 
    f_0(c) +
    f_\text{I} (s) +
    f_\text{I} (p)
    + \chi_s \, c \, s + \chi_p \, c \, p
    \big] 
    \, ,
\label{eq:free_energy}
\end{equation}
where $d$ refers to the number of spatial dimensions.
Note that, in general, the interaction between different species can be more intricate than the current bilinear form, e.g., when substrate and product concentrations are large enough for their virial coefficients to become relevant~\cite{Henninger2021}. 

\begin{figure}[tb]
\includegraphics{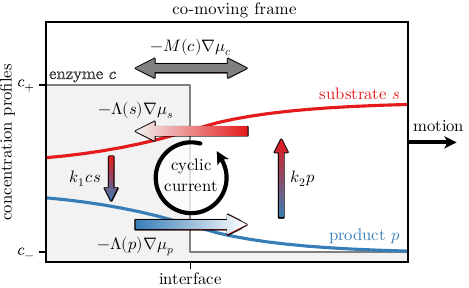}
\caption{%
\label{fig:model_sketch}%
Sketch of the model.
Gradients in chemical potential correspond to thermodynamic forces which give rise to average drift velocities and hence particle currents, as described by Eq.~\eqref{eq:constitutive}.
For enzymes, these currents are driven by purely passive enzyme-enzyme interactions, as well as interactions with substrates and products.
Energy is injected into the system via reactions which generate substrate and product concentration gradients and fluxes.
This motif can give rise to self-propelling states.
}
\end{figure}

When a particle of species ${ \varrho \in \{ s, p, c \} }$ is added or removed at a specific location $\mathbf{x}$, the system's free energy incurs thermodynamic costs given by the corresponding chemical potential ${\mu_\varrho = \delta\mathcal{F}/\delta \varrho}$.
In the framework of linear non-equilibrium thermodynamics~\cite{DeGroot2013, Balian2006}, gradients in chemical potentials act as thermodynamic forces, corresponding to conservative currents that drive particle exchange between adjacent points in space. 
To linear order, one has [Fig.~\ref{fig:model_sketch}]
\begin{equation} 
\begin{bmatrix}
    \mathbf{j}_s \\
    \mathbf{j}_p \\
    \mathbf{j}_c
\end{bmatrix} 
=
-
\begin{bmatrix}
    \Lambda(s) & 0 & 0 \\
    0 & \Lambda(p) & 0 \\
    0 & 0 & M(c)
\end{bmatrix} 
\cdot
\begin{bmatrix}
    \vec\nabla \mu_s \\ 
    \vec\nabla \mu_p \\
    \vec\nabla \mu_c
\end{bmatrix} \, ,
\label{eq:constitutive}
\end{equation}
where we have disregarded any cross-terms and considered the matrix of Onsager coefficients (mobilities) to be diagonal.
For simplicity, we assume that the mobilities $\Lambda (s)$ for the substrates and products share the same functional dependence, which in general is distinct from the  mobility $M(c)$ for the enzymes.
The thermodynamic equilibrium state of the system, defined by ${ \boldsymbol\nabla\mu_\varrho = 0 }$ for each ${ \varrho \in \{ s, p, c \} }$, is completely independent of the matrix of Onsager coefficients.
Hence, at this point, there are no constraints on the mobility functions yet.

In the following, we first study the dynamics of the enzymes, which is characterized by the continuity equation ${\partial_t c = - \boldsymbol\nabla \cdot \mathbf{j}_c}$. 
With the above expression for the enzyme current $\mathbf{j}_c$ this takes the form of a generalized Cahn-Hilliard (CH) model
\begin{equation}
    \partial_t c (\mathbf{x},t) 
    = 
    \boldsymbol\nabla \cdot 
    \bigl[ 
    M(c) \, \boldsymbol\nabla 
    \bigl( 
    \mu_0 (c) + \chi_s s + \chi_p p 
    \bigr) 
    \bigr]
    \, .
\label{eq:enzymes}
\end{equation}
Here, the chemical potential due to enzyme-enzyme interactions, ${\mu_0 (c) = - r (c-\tilde c) + u (c-\tilde c)^3 - \kappa \boldsymbol\nabla^2 c}$, is identical to the CH model~\cite{CahnHilliard1958, Cahn1961}.
For ${r > 0}$, the CH model shows phase separation into enzyme-rich and enzyme-poor regions, with concentrations ${c_\pm = \tilde c \pm \sqrt{r/u}}$.
In the following, consistently with our previous work~\cite{Demarchi2023}, we choose $c_+$ as unit of concentration, and ${\epsilon_0 \coloneqq r \, c_+}$ as unit of energy.
In contrast to the CH model, the enzyme currents in Eq.~\eqref{eq:enzymes} are not only driven by enzyme-enzyme interactions but also by concentration gradients in substrates and products.
We will next discuss mechanisms that can give rise to non-uniform substrate and product concentration profiles.

\subsection{Non-equilibrium reactions}
\label{sec:model_reactions}

In addition to the thermodynamic currents given by Eq.~\eqref{eq:constitutive}, biological systems also contain active processes that break detailed balance.
As in our prior work~\cite{Demarchi2023}, we consider a specific form of activity, whereby the enzymes $E$ catalyze a chemical reaction that turns substrates $S$ into products $P$ via a Michaelis-Menten-like kinetics, 
\begin{equation}
\label{eq:reaction_path_dissipate}
    \ce{$E + S$ <=>[$k_1$] $ES$ <=>> $EP$ <=> $E + P$} \, .
\end{equation}
For each step of this reaction kinetics, the framework of nonequilibrium thermodynamics relates the ratio between the forward and backward reaction rates to the change in free energy~\cite{ZwickerReview2022, Demarchi2023}.
To complete the description and determine the relative reaction rates of the different steps, one needs to invoke transition state theory and take into account the corresponding potential barriers that have to be overcome during a reaction~\cite{ZwickerReview2022, Demarchi2023}.
Here, we focus on a simplified scenario where the first step \ce{$E + S$ <=> $ES$} is rate-limiting (high potential barrier), and has no thermodynamically preferred direction (equal forward and backward rates).
In contrast, we assume that the non-rate-limiting second step \ce{$ES$ <=>> $EP$} is strongly biased towards products which have lower internal energy than substrates.
The overall reaction then proceeds with a net rate $k_1\,c\,s$ proportional to the concentration ${c \equiv [E]}$ of enzymes and the concentration ${s \equiv [S]}$ of substrates [Fig.~\ref{fig:model_sketch}].
In summary, the net reaction rate is determined by the first reaction step, while free energy is released in the second reaction step.

To maintain the dynamics and keep the system away from thermodynamic equilibrium\footnote{Alternatively, one could also consider a sufficiently large domain in which substrates are abundant, so that the system does not reach thermodynamic equilibrium on the time scales of interest.}, the free energy released from converting substrate into product needs to be resupplied by exchanging product with substrate, 
\begin{equation}
    \label{eq:reaction_path_resupply}
    \ce{$P$ + $F$ <=>>[$k_2$] $S$ + $W$} \, ,
\end{equation}
along a separate reaction path which consumes fuel $F$ and releases waste $W$.
We assume that the forward reaction of Eq.~\eqref{eq:reaction_path_resupply} is driven by an excessive abundance of (chemostatted) fuel, while the backward reaction is limited by negligible concentrations of (chemostatted) waste.
The mass action law then suggests a net reaction rate $k_2 \, p$ proportional to the concentration $p \equiv [P]$ of products [Fig.~\ref{fig:model_sketch}].
The rate coefficient $k_2$ is spatially uniform if the fuel molecules are very small and thus have a high diffusion coefficient, or if the large abundance of fuel saturates the binding kinetics of product and fuel.
Thus, as further discussed in Appendix~\ref{appendix:transition_state}, we consider a simplified scenario where the reaction rate $k_2$ is concentration-independent.

With the simplifications outlined above, the reaction-diffusion dynamics of substrates and products are given by [Fig.~\ref{fig:model_sketch}]
\begin{subequations}
\label{eq:system}
\begin{align}
\label{eq:substrates}
    \partial_t s 
    &= 
    \boldsymbol\nabla \cdot 
    \left( D \, \boldsymbol\nabla s + \Lambda \, s \, \chi_s \boldsymbol\nabla c  
    \right) - k_1 \, c \, s + k_2 \, p \,, \\
    \label{eq:products}
    \partial_t p 
    &= 
    \boldsymbol\nabla \cdot 
    \left( D \, \boldsymbol\nabla p + \Lambda \, p\, \chi_p \boldsymbol \nabla c  
    \right) + k_1 \, c \, s - k_2 \, p \,.
\end{align}
\end{subequations}
Here, to recover Fick's law of linear diffusion starting from the fluxes described by Eq.~\eqref{eq:constitutive} and the free energy Eq.~\eqref{eq:free_energy}, we have assumed ${\Lambda(s) = \Lambda \, s}$ and ${\Lambda(p) = \Lambda \, p}$ for the mobility functions, with constant mobility coefficient $\Lambda$.
The physical interpretation of this choice is that the chemical potentials $\mu_{\varrho}$, where ${\varrho \in \{ s,p \}}$, give rise to average drift velocities $\mathbf{v}_{\varrho} = - \Lambda \boldsymbol\nabla \mu_{\varrho}$.
For thermal Brownian motion of the particles, the diffusivity is set by the Einstein-Smoluchowsky relation ${D=\Lambda \, k_B T}$~\cite{Frey2005}.
Here, we take the liberty of deviating from this relation, as the system is considered to be driven away from thermal equilibrium by active processes that break detailed balance.
The decay rate of products sets a characteristic timescale for the reaction-diffusion dynamics, ${\tau_0 \coloneqq k_2^{-1}}$, while the corresponding diffusion length sets the length scale, ${l_0 \coloneqq \sqrt{D/k_2}}$.

\begin{figure}[tb]
\includegraphics{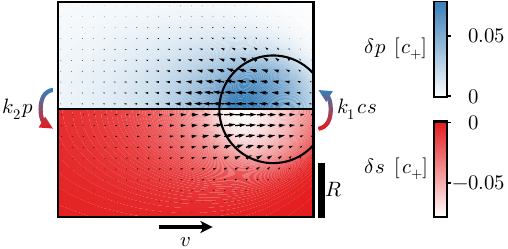}
\caption{%
\label{fig:moving_droplet}%
Concentration profiles of substrates and products for a 3D enzyme droplet which has radius $R$ (scale bar).
We here assume that the enzyme-rich region (with concentration $c_+$) has a sharp interface (black circle) with the enzyme-poor region (concentration $c_-$).
We obtain the steady-state concentration profiles by solving Eq.~\eqref{eq:system_comoving} numerically using FEM, for a scenario where the droplet moves with a prescribed constant speed ${v = 4.14 \, l_0 / \tau_0}$ through an open domain.
Colored arrows illustrate reactive fluxes: Enzymes catalyze the conversion of substrates into products with rate $k_1 \, c \, s$, while substrates are replenished from products with rate $k_2 \, p$.
Upper panel (blue) shows product concentration profile relative to the far-field, ${\delta p = p - p_\infty}$, while lower panel (red)  shows substrate concentration profile relative to the far-field, ${\delta s = s - s_\infty}$, both in units of the enzyme concentration $c_+$ within the droplet.
Vector field (thin black arrows) indicates the excess diffusive currents of substrates and products (a.u.) relative to the co-moving frame.
The remaining parameter values are ${c_-=0.5 c_+}$, ${R=l_0}$, ${k_1=2 k_2 / c_+}$, ${\Lambda=0}$, and ${s+p=c_+}$.
}
\end{figure}

We have previously shown that the coupled system of Eqs.~\eqref{eq:enzymes}~and~\eqref{eq:system} can give rise to self-propelling droplet states~\cite{Demarchi2023}.
The mechanism underlying this symmetry breaking is a feedback loop whereby droplet motion leads to asymmetric substrate and product concentration profiles, which drive enzyme currents, which in turn sustain droplet motion if enzyme-substrate interactions are more favorable than enzyme-product interactions.
We have previously shown that this nonequilibrium steady state can be accessed if the enzyme mobility is sufficiently high and if the reaction rates lie within some optimal range.
To better understand the underlying physics, in the present work, we focus on the analysis of droplets that exhibit a traveling steady state which moves with velocity $\mathbf{v}$.
These droplets are characterized by a traveling enzyme concentration profile ${c(\mathbf{x}-\mathbf{v} t)}$, which will be the focus of the next section.
Thus, we substitute the traveling wave ansatz ${s(\mathbf{x},t) = s(\mathbf{x}-\mathbf{v} t)}$ and ${p(\mathbf{x},t) = p(\mathbf{x}-\mathbf{v} t)}$ with the Galilean coordinate transformation ${\mathbf{z} \coloneqq \mathbf{x} - \mathbf{v} t}$.
In the corresponding co-moving frame, the steady-state concentrations of substrates and products are determined by 
\begin{subequations}
\label{eq:system_comoving}
\begin{align}
\label{eq:substrates_comoving}
    0 
    &= 
    \boldsymbol\nabla \cdot 
    \left( \mathbf{v} \, s + D \, \boldsymbol\nabla s + \Lambda \, s \, \chi_s \boldsymbol\nabla c  
    \right) - k_1 \, c \, s + k_2 \, p \,, \\
    \label{eq:products_comoving}
    0 
    &= 
    \boldsymbol\nabla \cdot 
    \left( \mathbf{v} \, p + D \, \boldsymbol\nabla p + \Lambda \, p\, \chi_p \boldsymbol \nabla c  
    \right) + k_1 \, c \, s - k_2 \, p \,,
\end{align}
\end{subequations}
where the gradient $\boldsymbol\nabla$ is taken with respect to $\mathbf{z}$.

Consider now a single round enzyme droplet characterized by a domain ${\mathcal{D} \coloneqq \{\mathbf{z} : |\mathbf{z}| < R \} }$ with high enzyme concentration ${c(\mathbf{z}) = c_+}$, which is surrounded by an enzyme-depleted domain (${|\mathbf{z}| \geq R}$) with concentration ${c(\mathbf{z}) = c_-}$.
Given such an enzyme concentration profile, we solved Eq.~\eqref{eq:system_comoving} by using finite element methods (FEM) implemented in FEniCSx~\cite{Logg2012, Alnaes2014, Scroggs2022a, Scroggs2022b, Baratta2023}.
To emulate the infinitely large open space through which the droplet moves, we imposed Dirichlet boundary conditions at the boundary of the finite simulation domain.
For the corresponding values of the far-field concentrations, we used the reactive equilibria ${s_\infty = n / (1+ k_1 c_\infty/k_2)}$ and ${p_\infty = n - s_\infty}$, where $n$ is the total average concentration of substrates and products and ${c_\infty \coloneqq \lim_{|\mathbf{z}|\to\infty} c(\mathbf{z})}$. 
In the present manuscript, since our main focus is not on condensate growth and dissolution, we neglect the supersaturation of the enzymes and thus make the approximation $c_\infty \approx c_-$.

The resulting concentration profiles and fluxes of substrates and products are shown in Fig.~\ref{fig:moving_droplet}.
Inside the condensate, where the enzyme concentration is high, substrates are converted into products.
Outside the condensate, products are restored to substrates.
These chemical reactions lead to concentration gradients which drive diffusive fluxes.
As indicated by the vector fields in Fig.~\ref{fig:moving_droplet}, products are eliminated from the droplet by diffusing out, while substrates are replenished in the droplet by diffusing in.
Taken together, these reactions and diffusion form a closed cycle of currents which maintains the non-equilibrium steady state.
Finally, because the condensate moves ballistically, all concentration fields are advected rearwards in the droplet frame of reference.
Due to this effect, the extrema of the substrate and product concentration profiles  trail behind the centroid of the condensate.
With increasing droplet speed, the maximum of the product concentration profile and the minimum of the substrate concentration profile move closer to the trailing edge of the condensate.
It is this concentration gradient of substrates and products which drives droplet motion.

\section{Self-consistency relation for driven condensate motion}
\label{sec:self_consistency_relation_generalized}

As a result of enzyme-mediated reactions, the concentration profiles of both substrates and products are in general non-uniform. 
These non-uniform distributions result in enzyme currents driven by enzyme-substrate and enzyme-product interactions, as described by the generalized CH equation [Eq.~\eqref{eq:enzymes}].
The associated thermodynamic force is given by
\begin{equation}
\label{eq:thermodynamic_force_density}
   \mathbf{f}(\mathbf{x},t) \coloneqq - \vec\nabla \big[ \chi_s \, s(\mathbf{x},t) + \chi_p \, p (\mathbf{x},t) \big] \, .
\end{equation}
While we are using this specific form here, the theory presented in the following is independent of the specific physical mechanism that generates the force field ${\mathbf{f}(\mathbf{x},t)}$.
The theory applies equally to externally imposed driving forces or to forces generated from active processes.
Given such an inhomogeneous forcing $\mathbf{f}(\mathbf{x},t)$, this section outlines how the resulting droplet velocity ${\mathbf{v}}$ can be determined self-consistently in arbitrary spatial dimensions.
This is a cornerstone of our work and generally applies to any system exhibiting phase separation into well-defined droplets, irrespective of the underlying free-energy functional describing the interactions causing phase separation.

\emph{Sharp interface limit.} 
The central concept underlying our theoretical analysis is a sharp interface limit, which assumes that the width of the droplet interface ${w = \sqrt{2\kappa/r}}$ is small compared to all other length scales relevant to the dynamics.
While this width diverges near the critical point (${r=0}$), it approaches a molecular length scale given by the characteristic range of protein-protein interactions when phase separation is sufficiently strong.
Therefore, we expect that a sharp interface limit serves as a good approximation when the condensates (droplets) are much larger in size than individual molecules.
Mathematically, we implement this limit by taking ${r \rightarrow \infty}$ and ${u \rightarrow \infty}$ while maintaining finite values for mesoscopic quantities such as the equilibrium enzyme concentrations, and the chemical potential for the enzymes. 
This chemical potential can be written in the form
\begin{equation}
    \mu_0 (c)
    = 
    r 
    \left[ - (c-\tilde c) + \frac{4}{\Delta c^2} (c-\tilde c)^3 - \frac{1}{2} w^2 \boldsymbol\nabla^2 c 
    \right] \, ,
    \label{eq:chemical_potential_c}
\end{equation}
where ${\Delta c \coloneqq c_+ - c_-}$ is the difference in the equilibrium enzyme concentrations between the inside and the outside of the droplet.
For the chemical potential to remain finite and thus physical in the asymptotic limit ${r \to \infty}$, the expression in the square brackets must vanish.
This is precisely the same condition as that of a CH theory, resulting in a $\tanh$-profile with width $w$, height $\Delta c$, and asymptotic values ${c_\pm=\tilde c \pm \sqrt{r/u}}$~\cite{CahnHilliard1958, Cahn1961}.
In the sharp interface limit we are considering here, this profile can be considered as piecewise constant. 

If there were deviations $\delta c$ around these equilibrium concentrations $c_\pm$, they would quickly relax back to equilibrium because the corresponding chemical potential, 
\begin{equation}
    \mu_0 = 2r \, \delta c - \frac{1}{2} \,r \, w^2 \boldsymbol\nabla^2 \delta c + \mathcal{O}(\delta c^2) \, ,
\end{equation}
would exhibit large gradients.
These chemical potential gradients, with ${|\vec\nabla \mu_0| \gg |\chi_s \vec\nabla s| + |\chi_p \vec\nabla p|}$, would transiently dominate the dynamics and drive currents that promptly restore a piecewise constant enzyme concentration profile.
As a result, all variations in the concentration profile are localized at the droplet boundary, where the competing terms in Eq.~\eqref{eq:chemical_potential_c} are all of the same order and much larger than the interaction terms with substrates and products, ${r \,\Delta c \gg |\chi_s \, s| + |\chi_p \, p|}$.

\emph{Approximation of round droplet shape.}
Next, we discuss the geometric shape of the droplet, which will determine the piecewise constant enzyme concentration profile in the sharp interface limit.
To do so, it is important to note that the effective surface tension of a droplet in the CH model is given by~\cite{Bray1994}
\begin{equation}
\label{eq:condensate_surface_tension}
    \gamma = \frac{1}{6} (\Delta c)^2 \sqrt{2 \kappa r} 
    = \frac{1}{6} (\Delta c)^2 r \, w \, ,
\end{equation}
with ${\Delta c = c_+ - c_-}$ the density jump at the interface.
Since the interface width $w$ is bounded from below by the molecule diameter, the effective tension diverges in the limit ${r \rightarrow \infty}$.
For this reason, we restrict ourselves to round droplets with a constant interface curvature.

\subsection{Sharp interface theory}

As the above considerations show, Eq.~\eqref{eq:chemical_potential_c} is no longer a mathematically well-defined expression for the chemical potential. 
Therefore, we take an alternative approach and consider the chemical potential as an unknown.
More specifically, it is a Lagrange multiplier field that enforces the condition of a piecewise constant concentration profile: ${c(\mathbf{x}) = c_+}$ within the domain $\mathcal{D}$ of the droplet and ${c(\mathbf{x}) = c_-}$ outside.
The following analysis will describe how to obtain an equation that determines the chemical potential.

\emph{Traveling wave solution.}
We first seek a traveling wave solution for the enzyme dynamics by substituting the ansatz ${c(\mathbf{x},t) = c(\mathbf{x}-\mathbf{v} t)}$ into Eq.~\eqref{eq:enzymes}. 
One finds that the steady state of the enzyme concentration profile (in the co-moving frame) is given by a balance condition between advective terms (left-hand side) and driving forces (right-hand side)
\begin{equation}
    -\mathbf{v} \cdot \boldsymbol\nabla c(\mathbf{z})
    = 
    \boldsymbol\nabla \cdot 
    \bigl[ 
    M(c) \, 
    \bigl(
    \boldsymbol\nabla 
    \mu_0(\mathbf{z}) 
    - \mathbf{f}(\mathbf{z})
    \bigr)
    \bigr]
    \, .
\label{eq:enzymes_comoving}
\end{equation}
As explained in the previous section, there is no explicit form for the chemical potential $\mu_0 (\mathbf{z})$ in the sharp interface limit. 
Instead, it serves as a Lagrange multiplier field that ensures constant enzyme densities inside and outside the droplet. 
This means that the enzyme concentration profile denoted $c(\mathbf{z})$ and the associated mobility function, $M(c)$, are assumed to be known.
Furthermore, we temporarily assume knowledge of the droplet velocity $\mathbf{v}$ and will later demonstrate how to determine it self-consistently.
Given the force field $\mathbf{f}(\mathbf{z})$, and with appropriate boundary conditions which will be discussed in the next section, one can now use FEM to determine the chemical potential---the only unknown in the present scheme.

In the sharp interface limit, solving Eq.~\eqref{eq:enzymes_comoving} requires a careful analysis since gradients in the enzyme concentration become singular at the interface. 
To address this, we rearrange terms in Eq.~\eqref{eq:enzymes_comoving} to express it as: 
\begin{subequations}    
\begin{equation}
    \boldsymbol\nabla \cdot \mathbf{J}_\Delta = 0 \, ,
\label{eq:effective_euler-b}
\end{equation}
with the current given by
\begin{equation}
    \mathbf{J}_\Delta 
    = 
    M(c) 
    \bigl(- \boldsymbol\nabla \mu_0 + \mathbf{f} 
    \bigr) 
    - 
    \mathbf{v} \, 
    \bigl( 
    c - c_\infty 
    \bigr) \, .
\label{eq:effective_euler-a}
\end{equation}
\end{subequations}
Equation~\eqref{eq:effective_euler-a} can be viewed as the integral of Eq.~\eqref{eq:enzymes_comoving} and, as explained in the following, interpreted as the divergence-free current of the enzymes in the reference frame of the droplet.
The first term describes the currents that arise in the laboratory frame due to gradients in the chemical potential of the enzymes and the given force field $\mathbf{f}(\mathbf{x})$.
The second term is an apparent advective enzyme current, which an observer in the co-moving frame will see due to the motion relative to the laboratory frame.
To ensure this interpretation we added the integration constant $\mathbf{v} \, c_\infty$ to Eq.~\eqref{eq:effective_euler-a}.
Then, the enzyme currents vanish far away from the droplet, ${\lim_{|\mathbf{z}|\to\infty}\mathbf{J}_\Delta(\mathbf{z}) = 0}$, where the enzyme concentration has a low value ${\lim_{|\mathbf{z}|\to\infty}c(\mathbf{z}) = c_\infty \approx c_-}$.
To approximate these far-field conditions in our numerical implementation, we use a finite domain with a size much larger than the droplet radius, and impose no-flux boundary conditions at the domain boundary.
This defines the boundary conditions for the calculation of the chemical potential.

\begin{figure*}[t]
\centering
\includegraphics{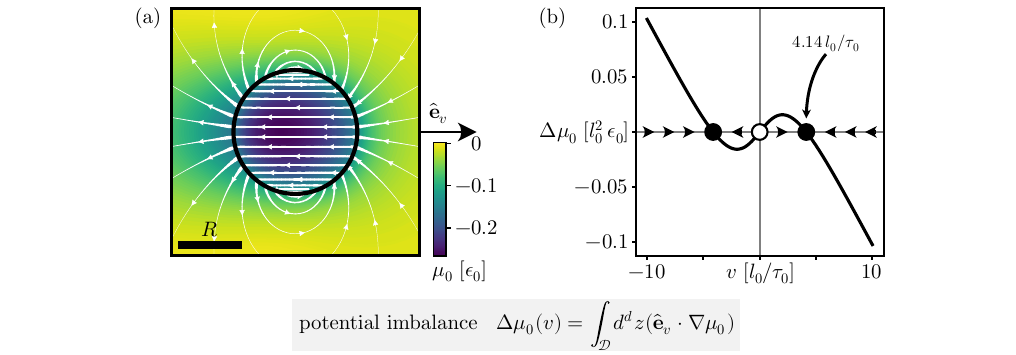}
\caption{%
\label{fig:self_consistency_example}%
\textbf{Sharp interface theory for a moving 3D enzyme droplet}.
\textbf{(a)} Chemical potential profile in the reference frame of a moving droplet (circle), obtained by solving Eq.~\eqref{eq:weak_form}.
The streamplot shows the (mesoscopic) enzyme velocity (a.u.) relative to the co-moving frame, $\mathbf{J}_\Delta / c$, as calculated from Eq.~\eqref{eq:effective_euler-a}. 
These currents are driven by a thermodynamic force, Eq.~\eqref{eq:thermodynamic_force_density}, due to interactions with substrates and products.
The droplet speed, reaction-diffusion dynamics, as well as substrate and product concentration profiles are taken from Fig.~\ref{fig:moving_droplet}.
Remaining parameters: ${\Delta\chi\coloneqq \chi_p - \chi_s = 4 r}$ and ${M(c) = M c}$ with ${M=10 D / \epsilon_0}$.
\textbf{(b)} Chemical potential imbalance $\Delta\mu_0(v)$ as a function of the droplet speed $v$ with stable (filled circles) and unstable (open circles) steady states. 
The effective phase space flow of the droplet speed $v$ driven by the chemical potential imbalance is shown as arrows.
For the indicated stable steady state, the corresponding enzyme chemical potential $\mu_0$ and flow field are shown in panel a), whereas the substrate and product concentration and flow fields are shown in Fig.~\ref{fig:moving_droplet}.
}
\end{figure*}

To now apply FEM, we convert Eq.~\eqref{eq:effective_euler-b} into an optimization problem (the \emph{weak form}) by multiplying by a test function $\phi(\mathbf{z})$ and integrating over the entire domain. 
In doing so, the singularity in the concentration profile can be lifted through integration by parts.
The numerical solution for the chemical potential profile then satisfies
\begin{equation}
    \int\! d^d z \, \bigl[\mathbf{J}_\Delta \cdot \boldsymbol\nabla\phi \bigr] = 0 
    \, .
\label{eq:weak_form}
\end{equation}
Solving this equation with the open-source FEM framework FEniCSx~\cite{Logg2012, Alnaes2014, Scroggs2022a, Scroggs2022b, Baratta2023}, one obtains the Lagrange multiplier field $\mu_0(\mathbf{z})$ which best enforces incompressibility of the enzyme currents [Eq.~\eqref{eq:effective_euler-b}] and thus maintains the enzyme concentration profile.
Finally, the corresponding enzyme currents can be obtained by inserting the numerical solution for the chemical potential into Eq.~\eqref{eq:effective_euler-a}.
Figure~\ref{fig:self_consistency_example}(a) shows both the chemical potential profile and the corresponding enzyme currents for a droplet in a self-propelling steady state, where the force field ${\mathbf{f}(\mathbf{z}) \coloneqq - \vec\nabla \big[ \chi_s \, s(\mathbf{z}) + \chi_p \, p (\mathbf{z}) \big]}$, cf. Eq.~\eqref{eq:thermodynamic_force_density}, is generated by inhomogeneous substrate and product concentration profiles as discussed in Sec.~\ref{sec:model_reactions}. 

In summary, we have developed an \emph{implicit adiabatic elimination scheme}, which complements standard (that is, explicit) methods.
Explicit methods consider how a given chemical potential profile [Eq.~\eqref{eq:chemical_potential_c}] gradually drives the system toward a state of phase separation.
This requires a dynamical resolution of the concentration profiles on the length scale of the interface.
In contrast, our approach starts with the premise that the system undergoes phase separation and treats the corresponding concentration profile as a fast degree of freedom.
This allows us to invert the logical flow, and instead analyze the chemical potential profile which is required to adiabatically maintain the concentration profile in the predefined configuration.
As a consequence of the reversal of the logical flow, prompted by the assumption of a sharp interface, explicit expressions for the chemical potential and the corresponding free energy of enzyme-enzyme interactions are abandoned in favor of an implicit description.
This implicit description [Eq.~\eqref{eq:weak_form}] is linear as a function of the chemical potential profile (now an unknown). 
It can therefore be solved several orders of magnitude faster than explicitly calculating the nonlinear dynamics of the concentration profiles in our full FEM simulations.

Importantly, the details of the original explicit model are not important for the reasoning underlying our model reduction scheme.
Instead, the dynamics is universally determined by the force field, the concentration (order parameter) profile, and the profile velocity.
We thus expect that our approach can also be adapted to Flory-Huggins-like models in the presence of non-equilibrium processes or to fully non-equilibrium systems exhibiting moving fronts, such as population expansion or reaction-diffusion dynamics.

\emph{A thermodynamic consistency relation for the chemical potential profile yields the droplet velocity.}
So far we have regarded the droplet velocity as a given quantity. 
We will now discuss how it can be determined through a condition on the chemical potential profile required by the thermodynamic consistency of the steady state. 

The sharp interface theory outlined above predicts the chemical potential field $\mu_0(\mathbf{z})$ and the associated enzyme currents in response to a specified force field $\mathbf{f}(\mathbf{z})$ for any given velocity $\mathbf{v}$.
While the force field $\mathbf{f}(\mathbf{z})$ originates from a non-equilibrium process or is imposed externally, gradients in the chemical potential are thermodynamic driving forces. 
Consequently, these gradients are linked to physical conditions that could help in determining the droplet velocity $\mathbf{v}$.
To characterize self-propelling and thus polar droplets, we define the chemical potential imbalance as a function of the droplet speed,
\begin{equation}
\label{eq:chemical_potential_imbalance}
    \Delta\mu_0(v) \coloneqq \int_{\mathcal{D}} \! d^d z \, \bigl( \mathbf{\hat{e}}_v \cdot \vec\nabla \mu_0 \bigr) \, .
\end{equation}
This definition is motivated by the fact that it gauges the overall thermodynamic driving force acting on a droplet: 
It gives the average chemical potential gradient within the droplet taking into account rotational symmetry about the direction of motion ($\mathbf{\hat{e}}_v$).
The dependence of $\Delta\mu_0(v)$ on the droplet speed is shown in Fig.~\ref{fig:self_consistency_example}(b), where we use for specificity a force field $\mathbf{f}(\mathbf{z})$ generated by inhomogeneous substrate and product concentration profiles, as described in section~\ref{sec:model_reactions}.

One may now interpret the function $\Delta\mu_0(v)$ as an effective force or its integral as an effective potential acting on the droplet speed $v$, closely resembling the dynamics of a one-component nonlinear system.
This indicates that the droplet speed at a steady state is determined by the condition that the chemical potential imbalance must vanish:
\begin{equation}
\label{eq:chemical_potential_imbalance-optimal}
    \Delta\mu_0(v) = 0 \, .
\end{equation}
In addition, the slope of the chemical potential disequilibrium provides information about the stability of the steady state: 
Positive values of $\Delta\mu_0(v)$ indicate that an additional force pointing from the leading to the trailing edge of the droplet is required to maintain a given velocity.
In the absence of this force, the droplet tends to speed up.
In contrast, for negative values of $\Delta\mu_0(v)$, the droplet tends to slow down.
Hence, a negative slope of $\Delta\mu_0(v)$ indicates stability. 
The resulting phase space flow is represented by the arrows in Fig.~\ref{fig:self_consistency_example}(b). 
In the following section, the above heuristic considerations are substantiated by detailed thermodynamic arguments.

\subsection{Thermodynamically consistent chemical potential is a consequence of energy conservation}
\label{sec:chemical_potential_imbalance_phase_portrait_analysis}
We will now illustrate the physical meaning of the thermodynamic consistency criterion [Eq.~\eqref{eq:chemical_potential_imbalance-optimal}], by linking our analysis to non-equilibrium thermodynamics.

\emph{Power dissipation.}
Using the sharp interface theory, we have determined the chemical potential profile $\mu_0(\mathbf{z})$ which is required to maintain a preset enzyme concentration profile $c(\mathbf{z})$, and a given droplet velocity $\mathbf{v}$.
In turn, in the presence of such a non-uniform chemical potential\footnote{For the Cahn-Hilliard model with a diffuse interface, the mechanisms giving rise to this chemical potential are encoded in the free energy density [Eq.~\eqref{eq:freeenergy_cahnhilliard}].}, variations in the concentration profile, e.g., due to droplet motion, translate into changes in the free energy ${\mathcal{F}_0[c] \coloneqq \int \! d^d x \, f_0(c)}$ associated with enzyme-enzyme interactions, 
\begin{equation}
    \label{eq:pseudo_free_energy_variational}
    \delta \mathcal{F}_0[c] = \int \! d^d x \, \mu_0(\mathbf{x}) \, \delta c(\mathbf{x}) \, .
\end{equation}
Thus, the chemical potential will dissipate free energy with a rate
\begin{equation}
    \partial_t \mathcal{F}_0[c] = \int \! d^d x \, \mu_0(\mathbf{x}) \, \partial_t c(\mathbf{x},t) \, .
\end{equation}
One can rewrite this expression by substituting the traveling wave ansatz, ${c(\mathbf{x},t) = c(\mathbf{x}-\mathbf{v} t)}$, and by partial integration, which yields
\begin{equation}
    \label{eq:pseudo_free_energy_flux}
    \partial_t \mathcal{F}_0[c] = \mathbf{v} \cdot \int \! d^d z \, c(\mathbf{z}) \, \boldsymbol\nabla\mu_0(\mathbf{z}) \, .
\end{equation}
Note that this is simply minus the power dissipated by a flux down a chemical potential gradient, which can also be derived by using fundamental relations between power, work and force\footnote{
The motion of particles, with net current ${\mathbf{j}(\mathbf{z}) = c(\mathbf{z}) \, \mathbf{v}(\mathbf{z})}$, in the presence of a force field $\mathbf{f}(\mathbf{z})$ will, over time, perform work with rate
\begin{equation*}
    \partial_t W = \int \! d^d z \, \mathbf{j}(\mathbf{z}) \cdot \, \mathbf{f}(\mathbf{z}) \, .
\end{equation*}
This general relation between power dissipation and force is independent of invoking thermodynamic arguments. 
One recovers Eq.~\eqref{eq:pseudo_free_energy_flux} by identifying $-\boldsymbol\nabla\mu_0$ as a thermodynamic force which extracts the work ${\delta W = -\delta\mathcal{F}_0}$ from the free energy functional.
}.

Utilizing that the  enzyme concentration profile 
$c(\mathbf{z})$ is piecewise constant, we partition the aforementioned integral into two domains: 
$\mathcal{D}$ within the droplet, and $\Omega\backslash\mathcal{D}$ outside of it, where $\Omega$ denotes the whole domain.
With ${c(\mathbf{z}) = c_+ \, \forall \, \mathbf{z}\in \mathcal{D}}$ and ${c(\mathbf{z}) = c_- \, \forall \, \mathbf{z}\notin \mathcal{D}}$, one then has
\begin{equation}
\label{eq:pseudo_free_energy_split_domain}
    \partial_t \mathcal{F}_0 = 
    \mathbf{v} \cdot \biggl[ 
    c_+ \int_{\mathcal{D}} \! d^d z \, \boldsymbol\nabla\mu_0 +
    c_- \int_{\Omega\backslash\mathcal{D}} \! d^d z \, \boldsymbol\nabla\mu_0 \biggr] .
\end{equation}
To reformulate the second term in the square brackets, we use the Gauss theorem which implies that the integral of $\boldsymbol\nabla\mu_0$ over the whole domain $\Omega$ vanishes
\begin{equation}
\label{eq:integral_split_domain}
    \int_{\Omega\backslash\mathcal{D}} d^d z \, \boldsymbol\nabla\mu_0 + \int_{\mathcal{D}} d^d z \, \boldsymbol\nabla\mu_0 = \oint_{\partial\Omega} \! d^{d-1}\boldsymbol{S} \, \mu_0 = 0 \, .
\end{equation}
Note that the last equality holds because the chemical potential must be constant in the far field and because the integral of the normal vector over a closed surface is always zero.
Substituting Eq.~\eqref{eq:integral_split_domain} into Eq.~\eqref{eq:pseudo_free_energy_split_domain}, one has
\begin{equation}
    \partial_t \mathcal{F}_0 = 
    v \, \Delta c \,
    \left[ \mathbf{\hat{e}}_v \cdot
    \int_{\mathcal{D}} \! d^d z \, \boldsymbol\nabla\mu_0 
    \right] ,
\end{equation}
where we have used $\mathbf{v} = v \, \mathbf{\hat{e}}_v$ and $\Delta c = c_+ - c_-$.
By comparing this expression with the definition of the chemical potential imbalance [Eq.~\eqref{eq:chemical_potential_imbalance}], one finds
\begin{equation}
\label{eq:power_dissipation}
    \partial_t \mathcal{F}_0 = v \, \Delta c \, \Delta \mu_0(v) \, .
\end{equation}
Therefore, the heuristically derived thermodynamic consistency relation ${\Delta\mu_0(v) = 0}$ [Eq.~\eqref{eq:chemical_potential_imbalance-optimal}] equates to the minimization of the free energy $\mathcal{F}_0[c]$ associated with enzyme-enzyme interactions.
In other words, in steady state, the currents along the chemical potential profile $\mu_0(\mathbf{z})$, which models enzyme-enzyme interactions, should cease to dissipate power.
In contrast, all of the power is dissipated by currents along the force field ${\mathbf{f}(\mathbf{z}) \coloneqq - \vec\nabla \big[ \chi_s \, s(\mathbf{z}) + \chi_p \, p (\mathbf{z}) \big]}$, cf. Eq.~\eqref{eq:thermodynamic_force_density}, where maintaining the substrate and product concentration profiles requires fuel.
In agreement with these arguments, we have confirmed that the chemical potential imbalance indeed vanishes in our simulations [Fig.~\ref{fig:self_consistency_test_comparison}(c)].

Note that, when the internal energy $\mathcal{U}$ of a system component does not change over time because it is in a (non-equilibrium) steady state, one can use the thermodynamic relation ${\mathcal{F} = \mathcal{U} - T \, \mathcal{S}}$ to relate Eq.~\eqref{eq:power_dissipation} to an entropy production rate, ${\partial_t \mathcal{S} = - T^{-1} \partial_t \mathcal{F}}$. One can use these concepts to also calculate the entropy production rate due to flows along the force field $\mathbf{f}(\mathbf{z})$.

\emph{Thermodynamic housekeeping.}
To elucidate the significance of the above result, we start with a scenario in which the force field is absent, ${\mathbf{f}(\mathbf{z}) = 0}$. 
Because there is no net driving force, thermodynamics requires that the droplet can only be in a stable steady state when it is at rest, ${v = v^\star = 0}$.
This state is characterized by a vanishing chemical potential imbalance, $\Delta \mu_0(v^\star) = 0$, and by a vanishing power dissipation.
What are then the characteristics of a hypothetical state with finite velocity?
To answer this question, we determined the chemical potential for a given droplet velocity by solving Eq.~\eqref{eq:weak_form}; for an illustration see Supplemental~Video~1. 
This shows that for any state with ${v \neq 0}$, the chemical potential at the trailing edge of the droplet must be higher than at its leading edge, so that ${v \, \Delta\mu_0(v) < 0}$, and the power dissipation $-\partial_t \mathcal{F}_0$ is positive [Eq.~\eqref{eq:power_dissipation}].
Hence, droplet motion would continuously siphon energy from the enzyme-enzyme interactions.
To then, in turn, adiabatically maintain enzyme-enzyme interactions and the enzyme concentration profile, one would need to inject energy into the system.
Taken together, energy conservation rules out a self-propelling droplet state in the absence of an external power supply.

To pursue these thermodynamic arguments further, we next analyze how the power dissipation changes when the steady state is perturbed.
To that end, we first evaluate
\begin{equation}
\label{eq:stability_first_order}
    \left. \frac{\partial}{\partial v} \, \partial_t \mathcal{F}_0 \right\rvert_{v = v^\star} = v^\star \Delta c \, \partial_v \Delta\mu_0 (v^\star) \, ,
\end{equation}
which vanishes for ${v^\star = 0}$.
Thus, the steady state is marginally stable at the linear level and one needs the second derivative to determine its stability: 
\begin{equation}
\label{eq:stability_second_order}
    \left. \frac{\partial^2}{\partial v^2} \, \partial_t \mathcal{F}_0 \right\rvert_{v = v^\star = 0} = 2 \Delta c \, \partial_v \Delta\mu_0 (v^\star) \biggr\rvert_{v^\star = 0} \, .
\end{equation}
Based on the arguments laid out in the previous paragraph, the steady state, which has vanishing power dissipation, is \emph{stable} when the power dissipation ${-\partial_t \mathcal{F}_0}$ is \emph{minimal} and hence Eq.~\eqref{eq:stability_second_order} is negative.
This leads us to investigate the implications of a local \emph{maximum} in the power dissipation as a function of the velocity.

To that end, we next consider a scenario in which a force field is present. 
Specifically, we consider a system where this force field is generated by the non-uniform concentration profiles of substrates and products [Eq.~\eqref{eq:thermodynamic_force_density}], which arise from reactions, diffusion, and advection in the co-moving frame of the condensate [Eq.~\eqref{eq:system_comoving}].
As in the above analysis, we again solve Eq.~\eqref{eq:weak_form} for given droplet velocities and monitor the chemical potential profile; see Supplemental~Video~2.
Importantly, since the force field depends on the droplet velocity, the chemical potential imbalance becomes a non-monotonic function of the droplet speed $v$; shown in Fig.~\ref{fig:self_consistency_example}(b) for the setup discussed here.
In particular, there are now several values $v^\star$ for the velocity at which the power dissipation and the chemical potential imbalance vanish, ${\Delta \mu_0 (v^\star) = 0}$.
As shown above in Eq.~\eqref{eq:power_dissipation}, the power dissipation $-\partial_t \mathcal{F}_0$ at ${v^\star = 0}$ still vanishes but now corresponds to a \emph{local maximum} as a function of the droplet velocity [cf. Fig.~\ref{fig:self_consistency_example}(b) and Eq.~\eqref{eq:stability_second_order}].
To maintain a state with a very small velocity, one would thus need to take energy out of the system, lest the condensate will spontaneously accelerate, which suggests that ${v^\star = 0}$ is an unstable steady state.
Next, we characterize the steady states ${v^\star \neq 0}$ and, without loss of generality, consider the case ${v > 0}$.
The slope of the chemical potential imbalance shown in Fig.~\ref{fig:self_consistency_example}(b) is negative, ${\partial_v \Delta\mu_0(v^\star) < 0}$. Together with Eq.~\eqref{eq:stability_first_order}, this implies that a further increase in the velocity would require constant energy injection.
Conversely, a slight decrease in the velocity would require constant energy elimination.
This is the signature of a stable steady state with a finite velocity.
These insights allow constructing a phase portrait [c.f.~arrows in Fig.~\ref{fig:self_consistency_example}(b)] reminiscent of dynamical systems theory~\cite{Strogatz2018}.

\subsection{Test of the sharp interface theory}
\label{sec:test_theory_nonreciprocal_case}

\begin{figure*}[htb]
\centering
\includegraphics{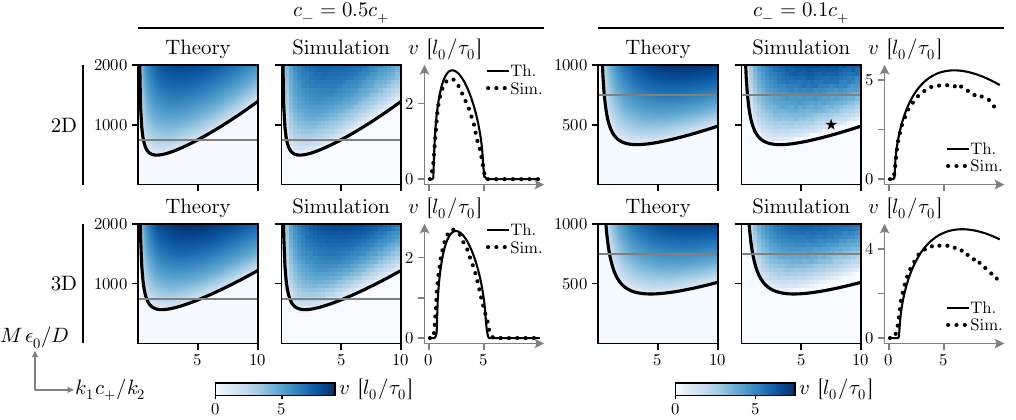}
\caption{%
\label{fig:self_consistency_test}
\textbf{Test of the sharp interface theory for moving enzyme condensates.}
We consider 2D and 3D droplets with two different concentrations in the dilute phase, ${c_- = 0.5 \, c_+}$ and ${c_- = 0.1 \, c_+}$, as indicated in the graph.
For each case, the first panel shows the predictions of the self-consistent sharp interface theory, while the second panel shows the results of the full FEM simulations solving Eqs.~\eqref{eq:enzymes}~and~\eqref{eq:system}.
In the third panel, we show a lineplot along the gray line indicated in the first two panels.
The predictions of the sharp interface theory for the phase boundary of the self-propulsion instability (solid lines) correctly mark the onset of droplet motion in the full FEM simulations, both for 2D and 3D systems.
Moreover, the theoretically predicted droplet velocities also agree quantitatively with the simulations. 
As parameters we used ${\chi_s = -0.05 r}$, ${\chi_p = -0.01 r}$, ${\Lambda=0}$, ${w = 0.1 l_0}$, ${R = l_0}$, and ${s+p = c_+}$.
In 2D, we simulated the full dynamics in a circular domain having radius ${L = 5 l_0}$ for ${c_- = 0.5 c_+}$ and radius ${L = 7 l_0}$ for ${c_- = 0.1 c_+}$. 
In 3D, we simulated the full dynamics in a cylindrical domain having radius ${L_r = 3 l_0}$ and length ${L_z = 10 l_0}$ for ${c_- = 0.5 c_+}$ and radius ${L_r = 4 l_0}$ and length ${L_z = 14 l_0}$ for ${c_- = 0.1 c_+}$.
The simulation data for ${c_- = 0.1 c_+}$ are taken from our previous work, Ref.~\cite{Demarchi2023}.
The exemplary set of parameters indicated by the star is further interrogated in Fig.~\ref{fig:self_consistency_test_comparison}.
}
\end{figure*}

To test our sharp interface theory, we compare its predictions, in particular the thermodynamic consistency condition [Eq.~\eqref{eq:chemical_potential_imbalance-optimal}], with numerical simulations of the full condensate dynamics that take into account the diffuse interface of the enzyme droplet as described by the generalized CH equation [Eq.~\eqref{eq:enzymes}].
In both approaches, the enzyme currents are driven by inhomogeneous substrate and product concentration profiles.
In the simulations of the full dynamics, we determine these concentration profiles by solving the time-dependent reaction-diffusion equations for substrates and products, which are given in Sec.~\ref{sec:model_reactions}, in the laboratory frame [Eq.~\eqref{eq:system}].
In our sharp interface theory, in the same way as above, we use the steady-state profiles [Fig.~\ref{fig:moving_droplet}] obtained by solving the corresponding reaction-diffusion-advection equations for substrates and products in the co-moving frame [Eq.~\eqref{eq:system_comoving}].
 
To determine the velocities where the thermodynamic consistency condition [Eq.~\eqref{eq:chemical_potential_imbalance-optimal}] is fulfilled, different numerical algorithms can be used. 
For instance, one could perform a parametric sweep of the chemical potential imbalance for different velocity values and directly identify the roots from the resulting graph, as shown in Fig.~\ref{fig:self_consistency_example}(b).
Here, as a numerically more efficient method, we use Newton iterations starting from an initial guess for the velocity; specifically, we set ${v = 10.0 \, l_0/\tau_0}$.
We find that the thermodynamic consistency condition correctly predicts the onset of the self-propulsion of spherical droplets [Fig.~\ref{fig:self_consistency_test}]. 
Moreover, as further discussed in Appendix~\ref{appendix:error}, the sharp interface theory reproduces the droplet speed with reasonable quantitative accuracy.
This holds true even for droplets with weak phase separation [${\Delta c \ll c_+}$, Fig.~\ref{fig:self_consistency_test}] and for those that dissolve over time due to enzyme-substrate and enzyme-product interactions [Supplemental Movies~3~and~4 and Appendix~\ref{appendix:initial_conditions}].

\subsection{Analytically solvable limiting case of the sharp interface theory}
\label{sec:analytical_solution_sharp_interface_theory}

\begin{figure*}[htb]
\centering
\includegraphics{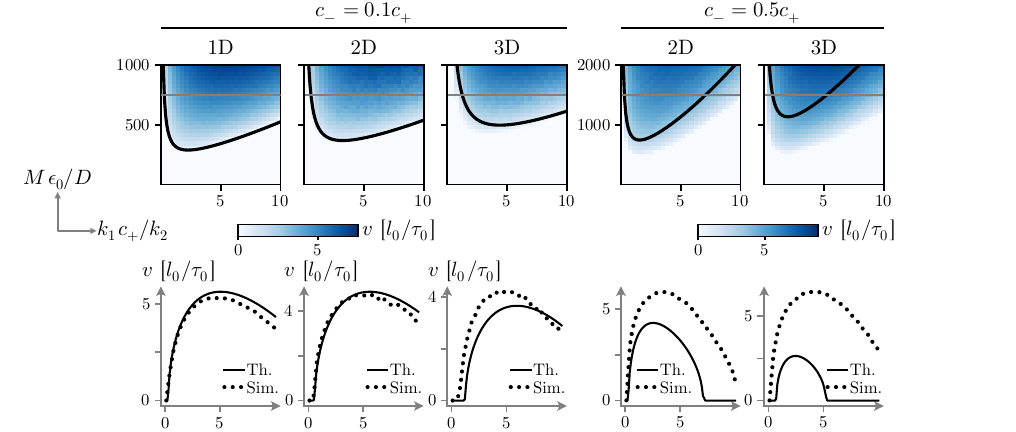}
\caption{
\label{fig:self_consistency-naive} 
\textbf{Naive approach of calculating the droplet velocity by neglecting all enzyme currents.}
The left group of panels corresponds to droplets with strong phase separation, ${c_- = 0.1 \, c_+}$, while the right group of panels indicates droplets with weak phase separation, ${c_- = 0.5 \, c_+}$.
Panels show droplet speed (color code) in the full FEM simulations; data and parameters are identical to Fig.~\ref{fig:self_consistency_test}.
Solid lines indicate naive theoretical prediction of the phase boundary of the self-propulsion instability, based on Eq.~\eqref{eq:effective_euler-a-simplified-final}.
The panels in the bottom row show a lineplot along the gray lines indicated in the top row.
This approximation gives acceptable results when phase separation is strong (${c_-/c_+ \sim 0}$, left group of panels), but fails in geometries whose dimension is larger than one if the enzymes are only weakly depleted from the low-concentration phase (${c_- \sim c_+}$, right group of panels).
}
\end{figure*}

The sharp interface analysis presented above typically requires a multi-step numerical evaluation, while closed analytical solutions exist only in special cases.
To illustrate the validity of our numerical approach beyond a comparison between simulation and theory, we will now discuss one such special case where an analytic solution can be found.
We consider a scenario where the enzyme flux $\mathbf{J}_\Delta$ vanishes outside the condensate.
This applies in three specific cases:
\begin{enumerate*}[label=(\roman*)]
\item For a 1D system, where the enzyme current must be spatially uniform due to Eq.~\eqref{eq:effective_euler-b}, the no-flux far-field condition implies that the current must vanish in the entire domain;
\item when the enzyme mobility vanishes in the low-concentration phase, ${M(c_-) = 0}$, it follows from Eq.~\eqref{eq:effective_euler-a} that there can be no flux;
\item when the enzymes are completely depleted in the low-concentration phase, ${c_- = 0}$, the lack of particles implies a vanishing net current.
\end{enumerate*}
If at least one of these conditions holds, one can use an approach similar to that described in our earlier work~\cite{Demarchi2023}. 
We first integrate Eq.~\eqref{eq:effective_euler-a} over the domain of the condensate:
\begin{equation}
    \int_{\mathcal{D}} \! d^d z \, \mathbf{J}_\Delta 
    = M(c_+) \int_{\mathcal{D}} \! d^d z \,
    \bigl(- \boldsymbol\nabla \mu_0 + \mathbf{f} 
    \bigr) 
    - V \, \Delta c \,
    \mathbf{v} \, ,
\label{eq:effective_euler-a-simplified}
\end{equation}
where ${V \coloneqq \int_{\mathcal{D}} \! d^d z}$ is the volume of the droplet. 
We will now show that the left-hand side of Eq.~\eqref{eq:effective_euler-a-simplified} is zero.
To that end, we manipulate the expression ${\int \! dV \, \boldsymbol\nabla\cdot(\mathbf{x}\otimes \mathbf{J}_\Delta)}$ to find
\begin{equation}
    \int \! dV \, \mathbf{J}_\Delta + \int dV \, \mathbf{x} \, (\nabla\cdot\mathbf{J}_\Delta) 
    = \oint d\mathbf{S} \cdot (\mathbf{x}\otimes\mathbf{J}_\Delta) \, ,
\end{equation}
where we applied the product rule on the left-hand side and the divergence theorem on the right-hand side.
Taking into account that the enzyme currents are divergence-free [Eq.~\eqref{eq:effective_euler-b}], and assuming that they vanish outside the condensate and thus along its boundary $\partial\mathcal{D}$, implies that the average current in the droplet must vanish. 
This is intuitive, because it simply means that the center of mass is stationary in the co-moving frame.
Equation~\eqref{eq:effective_euler-a-simplified} can be further simplified by using the thermodynamic consistency criterion for the chemical potential, Eq.~\eqref{eq:chemical_potential_imbalance-optimal}.
By projecting Eq.~\eqref{eq:effective_euler-a-simplified} on the axis $\mathbf{\hat{e}}_v$ and solving for the droplet speed $v$, one then finds
\begin{equation}
    v 
    = \frac{M(c_+)}{V \, \Delta c} \int_{\mathcal{D}} \! d^d z \, \mathbf{\hat{e}}_v \cdot \mathbf{f}(\mathbf{z}) 
    \, .
\label{eq:effective_euler-a-simplified-final}
\end{equation}
This is also what one would expect if every material point of the condensate is uniformly transported with the same velocity $\mathbf{v}$.
Then, the average driving force experienced by each material point is balanced by its viscous friction with the surrounding medium.

However, we would like to stress that this relation, Eq.~\eqref{eq:effective_euler-a-simplified-final} ceases to be valid when the enzyme currents are not uniform, such as in systems with more that one spatial dimension.
Specifically, we expect the error in Eq.~\eqref{eq:effective_euler-a-simplified-final} to become larger when phase separation is weak, ${\Delta c \ll c_+}$.
Figure~\ref{fig:self_consistency-naive} shows that naively using Eq.~\eqref{eq:effective_euler-a-simplified-final} to calculate the droplet velocity self-consistently in response to the concentration profiles of substrates and products, which themselves depend on the droplet velocity, is accurate only under specific conditions. 
For instance, it is applicable in scenarios like the 1D case discussed in Ref.~\cite{Demarchi2023}, or it can be a good approximation when the low concentration phase is almost depleted (${c_-/c_+ \approx 0}$). 

Interestingly, the simulations of the full dynamics show that self-propulsion starts much earlier than would be predicted based on Eq.~\eqref{eq:effective_euler-a-simplified-final}, which neglects the inhomogeneity in the enzyme currents.
Thus, liquid-like droplets can move much faster than one would expect if each of their material points were to be advected by a uniform velocity (that is, faster than solid-like condensates).
The reason for this is that the net movement of the condensate in terms of its concentration profile does not require any actual mass transport of the molecules in the highly concentrated phase over the same distance.
This effect can be attributed to the divergence-free circulation currents that transport enzymes in the co-moving frame of the condensate:
In the high-concentration phase, enzymes are transported from the leading edge of the droplet to the trailing edge.
The local outflow of enzymes causes the trailing edge of the condensate to retract.
Conversely, due to mass conservation, these enzymes return in the low-concentration phase from the trailing edge to the leading edge [Fig.~\ref{fig:self_consistency_example}(a)].
This influx of enzymes causes the leading edge of the condensate to expand.

\section{Reciprocal interactions control type of self-propulsion instability}
\label{sec:reciprocal_interactions}

\begin{figure}[t]
\centering
\includegraphics{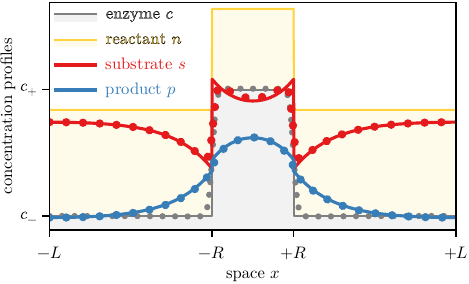}
\caption{
Stationary concentration profiles for enzymes (gray), substrates (red), and products (blue) of a single droplet in a finite-sized 1D domain $[-L,L]$. 
The total reactant (substrate and product) mass is shaded in yellow.
Dots correspond to FEM simulation results, while solid lines are analytical solutions for the substrate and product concentration profiles in the sharp interface limit (gray shading indicates droplet).
Attractive enzyme-substrate and enzyme-product interactions, ${\chi_s = -0.05 r}$, ${\chi_p = -0.01 r}$, in the presence of reciprocal interactions, ${\Lambda = 20 D/\epsilon_0}$, lead to a concentration jump in substrates and products.
The other parameters are given by ${c_- = 0.1 c_+}$, ${w = 0.1 l_0}$, ${R = l_0}$, ${L = 5 l_0}$, ${M = 100 D/\epsilon_0}$, ${k_1 = k_2/c_+}$, and ${\langle s+p \rangle = c_+}$.
}
\label{fig:stationary_profiles_reciprocal}
\end{figure}

In our previous work~\cite{Demarchi2023}, and in Figs.~\ref{fig:moving_droplet}-\ref{fig:self_consistency-naive} we have focused on the limiting case where interactions are weak and therefore the Flory-Huggins parameters $\chi_{s,p}$ are small. 
In this limit, the terms proportional to $\Lambda$ in the diffusion-reaction equations for the substrates and products, Eq.~\eqref{eq:system}, can be neglected, corresponding to a non-reciprocal limit.
In the following we will study the more general case ${\Lambda > 0}$ and the resulting consequences of the (partial) restoration of reciprocity.
To keep our analysis analytically tractable\footnote{We carry out the analytic calculations with the computer algebra system Mathematica~\cite{Mathematica} and provide the corresponding notebooks in Ref.~\cite{CodeGithub}.}, we again mostly return to a one-dimensional (1D) system.
As discussed in detail in Appendix~\ref{sec:substrate_interaction_profile_reciprocal}, we solve Eq.~\eqref{eq:system_comoving} to obtain the steady state substrate and product concentration profiles in the co-moving frame of the droplet.
An important feature of these concentration profiles is that reciprocal interactions, for $\Lambda > 0$, induce a concentration jump at the droplet interface [Fig.~\ref{fig:stationary_profiles_reciprocal}]:
\begin{equation}
    \frac{\varrho\vert_\text{in}}{\varrho\vert_\text{out}} 
    = 
    \exp \biggl[ -\frac{\Lambda \, \chi_\varrho \, \Delta c}{D} \biggr] \, ,  
\label{eq:concentration_jump_main}
\end{equation}
where ${\varrho \in \{ s,p \}}$.
This quantifies the meaning of ``weak interactions'', where the concentration jump is small, versus ``strong interactions'', where the concentration jump is large.

\subsection{Reciprocal interactions can inhibit droplet self-propulsion}
\label{sec:reciprocal_propulsion_instability}

\begin{figure}[t]
\centering
\includegraphics{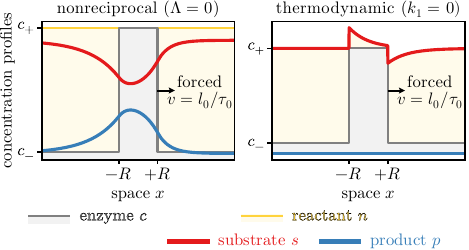}
\caption{
Concentration profiles for substrates (red), and products (blue) for a single droplet (gray) which is moving at a prescribed velocity through an open domain.
The total reactant (substrate and product) mass is shaded in yellow.
In the nonreciprocal case, substrates are depleted at the trailing edge of the condensate, while products are enriched there.
In contrast, in the thermodynamic limit, substrates are enriched at the trailing edge while products are not produced and are thereby absent from the system. 
Parameters unless specified otherwise: ${\chi_s = -0.05 r}$, ${\chi_p = -0.01 r}$, ${\Lambda = 4 D/\epsilon_0}$, ${c_- = 0.1 c_+}$, ${R = l_0}$, ${k_1 = k_2/c_+}$, and ${\langle s+p \rangle = c_+}$.
}
\label{fig:thermodynamic_vs_nonreciprocal}
\end{figure}

Substrates and products can be enriched inside of a droplet simply through attractive interactions with enzymes [Eq.~\eqref{eq:concentration_jump_main} and Appendix~\ref{sec:substrate_interaction_profile_reciprocal}], without relying on non-equilibrium reactions.
As shown in Fig.~\ref{fig:stationary_profiles_reciprocal}, this can lead to a scenario where the concentration of substrates is actually higher in the condensate than in the surrounding solution, despite being consumed by enzymatically catalyzed reactions.
To illustrate what this means for droplet self-propulsion, we consider an enzyme condensate which moves (or is pulled) with a prescribed velocity through an open domain.
Moreover, to disentangle the effects of reciprocal interactions from those of non-equilibrium reactions, we compare the thermodynamic limit where reactions are absent, ${k_1 = k_2 = 0}$ and ${\Lambda > 0}$, to the nonreciprocal limit, ${k_{1,2} > 0}$ and ${\Lambda = 0}$, cf.~Fig.~\ref{fig:thermodynamic_vs_nonreciprocal}. 
In the nonreciprocal limit, as shown in our previous work~\cite{Demarchi2023}, one finds that products are accumulated and substrates are depleted at the trailing edge of the droplet.
When enzyme-substrate attraction is stronger than enzyme-product attraction, this can push the droplet towards its leading edge and thereby drive self-propelled motion.
In contrast, in the thermodynamic limit, products are absent and attractive enzyme-substrate interactions cause accumulation of substrates at the trailing edge of the condensate.
Thus, the droplet will be pulled back towards its trailing edge, and condensate motion cannot be sustained.
Based on these competing mechanisms, we expect that reciprocal interactions can potentially inhibit reaction-induced droplet self-propulsion.

Having determined the substrate and product concentration profiles, we now quantify the effect of reciprocal interactions on the self-propulsion instability.
As we consider a 1D system, one could directly use Eq.~\eqref{eq:effective_euler-a-simplified-final} to obtain the droplet velocity in response to any inhomogeneous distribution of substrates and products, as we demonstrated recently~\cite{Demarchi2023},
\begin{equation}
\label{eq:self_consistency_relation_1D}
    v = -\frac{M c_+}{2 R \Delta c} \big[\chi_s \, \Delta s(v) + \chi_p \, \Delta p(v)\big] \, .
\end{equation}
Here $\Delta s(v)$ and $\Delta p(v)$ are the substrate and product concentration differences, respectively, between the right and the left boundary of the moving droplet.
However, from this approach it does not immediately become clear which of the self-consistent solutions are the stable attractors for the dynamics.
To close this gap in understanding, we follow a slightly different route by exploiting the framework presented in Sec.~\ref{sec:self_consistency_relation_generalized}, which not only yields the fixed points but also informs about their stability.

To that end, we compute the chemical potential imbalance $\Delta\mu_0(v)$, as defined by Eq.~\eqref{eq:chemical_potential_imbalance} and illustrated in Fig.~\ref{fig:self_consistency_example}(b) for a 3D droplet. 
For a 1D droplet, as discussed in Sec.~\ref{sec:analytical_solution_sharp_interface_theory}, our theory simplifies considerably because the divergence-free enzyme currents must vanish in the far-field and thus in the entire 1D domain, $J_\Delta = 0$.
By projecting Eq.~\eqref{eq:effective_euler-a-simplified} on the axis $\mathbf{\hat{e}}_v$, substituting the definition of the chemical potential imbalance [Eq.~\eqref{eq:chemical_potential_imbalance}] and solving for the latter, one finds
\begin{equation}
    \Delta\mu_0(v) 
    = \int_{\mathcal{D}} \! d z \,
    \bigl( \mathbf{\hat{e}}_v \cdot \mathbf{f} 
    \bigr) 
    - \frac{V \, \Delta c}{M(c_+)} \,
    v \, ,
\end{equation}
where ${\Delta c = c_+ - c_-}$ and ${V = 2R}$ is the droplet volume.
Finally, substituting the thermodynamic force field [Eq.~\eqref{eq:thermodynamic_force_density}] shows that the chemical potential imbalance for a 1D droplet is given by
\begin{equation}
\label{eq:chemical_potential_imbalance_1D}
    \Delta\mu_0(v) = - \chi_s \, \Delta s(v) - \chi_p \, \Delta p(v) - \frac{2 R \Delta c}{M c_+} \, v \, ,
\end{equation}
where we have assumed ${M(c_+) = M c_+}$ with constant $M$.
Note that Eqs.~\eqref{eq:chemical_potential_imbalance_1D}~and~\eqref{eq:self_consistency_relation_1D} are equivalent when the chemical potential imbalance vanishes, which corresponds to the fixed points of the dynamics.
As discussed in Sec.~\ref{sec:chemical_potential_imbalance_phase_portrait_analysis}, unstable (stable) fixed points are characterized by a positive (negative) slope of the chemical potential imbalance as a function of the velocity [Fig.~\ref{fig:self_consistency_example}(b)].
Thus, the chemical potential imbalance contains all information necessary to analyze the phase space flow.

\subsection{Discussion and test of the sharp interface theory}
\label{sec:discussion_and_test_reciprocal_theory}

\begin{figure*}[t]
\centering
\includegraphics{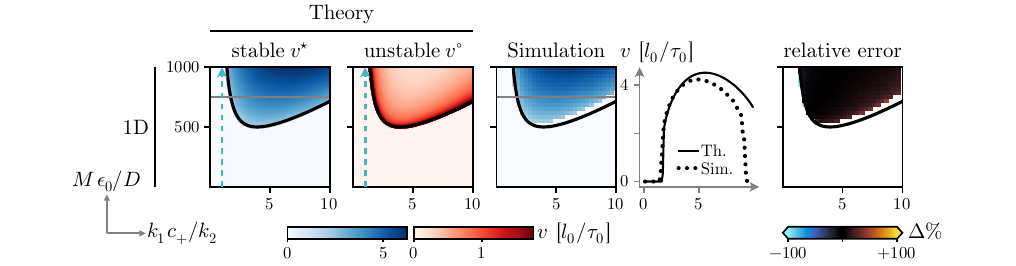}
\caption{Speed of a self-propelled 1D droplet as a function of the enzyme mobility $M$ and the reaction rate $k_1$.
The first panel shows the predicted droplet speed $v^\star$ of the stable steady state (blue color scale), while the second panel shows the unstable branch $v^\circ$ (red color scale), from the self-consistent sharp interface theory [Eqs.~\eqref{eq:chemical_potential_imbalance_1D}~and~\eqref{eq:chemical_potential_imbalance-optimal}].
The third panel shows the measured droplet speed $v_\text{sim}$ (blue color scale) in the full FEM simulations [Eqs.~\eqref{eq:enzymes}~and~\eqref{eq:system}].
The fourth panel shows a lineplot along the horizontal gray lines, to compare the prediction of our theory to the simulations.
The fifth panel shows the percentile error of the prediction, ${\Delta \coloneqq (v^\star - v_\text{sim})/v_\text{sim}}$ (blue-orange color scale).
Parameters: ${\chi_s = -0.05 \, r}$, ${\chi_p = -0.01 \, r}$, ${\Lambda = 4 D/\epsilon_0}$, ${c_- = 0.1 \, c_+}$, ${R = l_0}$, and ${\langle s+p \rangle = c_+}$.
The additional parameters required for the full FEM simulations are given by ${w = 0.1 \, l_0}$, and ${L = 30 \, l_0}$.
}
\label{fig:velocities_lambda}
\end{figure*}

One of the current limitations of the sharp-interface theory, which can be addressed in future studies as discussed in Appendix~\ref{appendix:ambiguity_concentration_difference}, is that it cannot resolve the concentration profiles inside the phase boundaries.
Hence, it is crucial to verify the predictions of the sharp interface theory against full FEM simulations, which feature dynamic and smooth concentration profiles of substrates and products [Eq.~\eqref{eq:system}] as well as enzymes [Eq.~\eqref{eq:enzymes}].
In this section, after quantitatively comparing theory and simulations, we also discuss central features of the bifurcation which leads to the onset of droplet self-propulsion.

To predict the speed of the biomolecular condensates with the sharp interface theory, we proceed as outlined in Sec.~\ref{sec:test_theory_nonreciprocal_case} for the nonreciprocal case.
First, we determine the chemical potential imbalance $\Delta\mu_0(v)$ [Eq.~\eqref{eq:chemical_potential_imbalance_1D}] as a function of the droplet velocity.
Then, we select the maximal droplet velocity $v^\star$ which satisfies the thermodynamic consistency criterion ${\Delta\mu_0(v^\star) = 0}$ [Eq.~\eqref{eq:chemical_potential_imbalance-optimal}].
For larger droplet velocities, the chemical potential imbalance is a monotonically decreasing and strictly negative function [Eq.~\eqref{eq:chemical_potential_imbalance_1D}],
\begin{equation}
    \Delta\mu_0(v) \approx - \frac{2 R \Delta c}{M c_+} \, (v - v^\star) \, .
\end{equation}
This implies ${\partial_v \Delta\mu_0(v^\star) < 0}$ and, following the discussion in Sec.~\ref{sec:chemical_potential_imbalance_phase_portrait_analysis}, that higher droplet velocities ${v>v^\star}$ would require additional energy input ${-\partial_t \mathcal{F}_0 > 0}$ [Eq.~\eqref{eq:power_dissipation}].
Therefore, we conclude that the velocity $v^\star$ must correspond to a stable steady state.
A more detailed discussion of the bifurcation diagram is deferred to a later point of this section.

As shown in Fig.~\ref{fig:velocities_lambda}, we find good agreement of the velocities and shape of the phase diagram predicted by the sharp interface theory with our full FEM simulations.
As in the nonreciprocal case, see Appendix~\ref{appendix:error}, the predictions become less accurate for large reaction rates, possibly because the diffusion length ${l_\pm = \sqrt{D/(k_1 c_\pm + k_2)}}$ becomes shorter compared to the width of the interface.
This parameter regime conflicts with the sharp interface limit, which requires that the interface width is small compared to all other relevant length scales.

Strikingly, unlike in the nonreciprocal case (${\Lambda = 0}$), in which the droplet velocity increased continuously from ${v = 0}$ upon reaching a critical enzyme mobility or reaction rate [Fig.~\ref{fig:self_consistency_test}], a sudden jump in the droplet speed from ${v = 0}$ to a finite value is observed for ${\Lambda > 0}$ [Fig.~\ref{fig:velocities_lambda}].
This points towards a subcritical bifurcation.
The bifurcation diagram of a subcritical transition is characterized by the stable branch ${v = v^\star}$, the trivial branch ${v = 0}$, which can change its stability depending on the parameters, and an unstable branch ${v = v^\circ}$ with ${0 \leq v^\circ \leq v^\star}$.
The binodal phase boundary is defined by the appearance of the stable branch ${v = v^\star}$, which is shown in the first panel of Fig.~\ref{fig:velocities_lambda}.
In contrast, the spinodal phase boundary is characterized by the disappearance of the unstable branch ${v = v^\circ}$ at the intersection ${v^\circ = 0}$.
This unstable branch is present within the parameter space examined in Fig.~\ref{fig:velocities_lambda}, as shown in the second panel, i.e., we are in the binodal regime.
To then observe droplet self-propulsion in this bistable regime between the binodal and the spinodal line, one needs to provide a sufficiently large perturbation.
To provide such a perturbation in our simulations, we initialized the concentration profiles of substrates and products with added noise.

\begin{figure}[tb]
\centering
\includegraphics{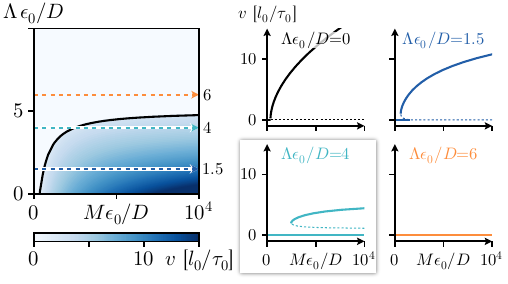}
\caption{%
Droplet speed $v^\star$ (blue color scale) predicted by the sharp interface theory, as a function of the enzyme mobility $M$ and the mobility $\Lambda$ of substrates and products in response to a gradient in enzymes. 
Bifurcation diagrams on the right correspond to slices along the dashed arrows in the left panel, for several selected values of $\Lambda$. 
Highlighted panel corresponds to slice along the dashed arrow in Fig.~\ref{fig:velocities_lambda}. 
Solid lines indicate stable branches, whereas dashed lines indicate branches of unstable fixed points.
Parameters: ${\chi_s = -0.05 \, r}$, ${\chi_p = -0.01 \, r}$, ${c_- = 0.1 c_+}$, ${R = l_0}$, ${k_1 = k_2/c_+}$, and ${\langle s+p \rangle = c_+}$.
}
\label{fig:self_propulsion_reciprocal}
\end{figure}

In the previous paragraph, we mapped out and discussed the phase diagram of the droplet velocity as a function of the enzyme mobility and reaction rates.
Next, we aim to better understand the role of the mobilities of substrates and products, ${\Lambda(s) = \Lambda s}$ and ${\Lambda(p) = \Lambda p}$.
To that end, we now keep the reaction rates fixed while varying the enzyme mobility $M$ and the reciprocity parameter $\Lambda$.
For cross-reference, a slice along the dashed arrows in the first two panels of Fig.~\ref{fig:velocities_lambda} corresponds to the bifurcation diagram depicted in the highlighted panel  in Fig.~\ref{fig:self_propulsion_reciprocal}.

First, we observe that the transition to self-propulsion is supercritical (continuous) for small substrate and product mobilities ${\Lambda = 0}$, as shown in the first panel on the right side of Fig.~\ref{fig:self_propulsion_reciprocal}.
For general ${\Lambda > 0}$, condensate self-propulsion sets in through a subcritical, imperfect pitchfork bifurcation.
Moreover, consistent with our arguments in the previous section, we observe that an increase in $\Lambda$ reduces the size of the parameter region where droplet self-propulsion can be observed.
Specifically, our findings indicate the existence of an upper bound for $\Lambda$, below which droplets display self-propulsion.
By using the Einstein-Smoluchowsky relation ${D=\Lambda \, k_B T}$ to relate the mobility of substrates and products to their diffusion coefficient in the case of thermal Brownian motion~\cite{Frey2005}, the vertical axis in Fig.~\ref{fig:self_propulsion_reciprocal} can be read as ${\Lambda \epsilon_0 / D = \epsilon_0 / (k_B T)}$.
This suggests that for the parameters used in Fig.~\ref{fig:self_propulsion_reciprocal}, the energy scale for the enzyme-substrate and enzyme-product interactions must be no more than ${5 k_B T}$ to observe droplet self-propulsion. 
The results in Fig.~\ref{fig:self_propulsion_reciprocal} also show that the mobility of substrates and products must be small compared to the mobility of enzymes.

Given these observations, what are necessary conditions to observe droplet self-propulsion for large substrate and product mobilities ${\Lambda}$ and for strong enzyme-substrate and enzyme-product interactions?
To answer this question, in the following we use the sharp interface theory to further elucidate how ${\Lambda}$ affects droplet self-propulsion and, moreover, when the transition is subcritical (discontinuous) or supercritical (continuous).

\subsection{Analysis of the role of reciprocity}
\label{sec:analysis_role_reciprocity}

Before we continue with our analysis, we first summarize a few key results that will be used in the following.
In our previous work~\cite{Demarchi2023} we established the mobility  of the enzymes $M$ as a key control parameter for observing  droplet self-propulsion.
We have found that in the absence of reciprocal interactions (${\Lambda = 0}$), self-propulsion always sets in if the mobility is sufficiently large, cf.~Fig.~\ref{fig:self_consistency_test} and Fig.~\ref{fig:self_propulsion_reciprocal}.
This can be understood from Eq.~\eqref{eq:chemical_potential_imbalance_1D}, which we rewrite as 
\begin{equation}
\label{eq:chemical_potential_imbalance_1D_rewritten}
    \Delta\mu_0(v) = 
    \Delta\mu^\star_0(v)
    - \frac{2 R \Delta c}{M c_+} \, v 
    \, .
\end{equation}
This chemical potential imbalance has an upper bound given by the limit $M \to \infty$, defining the maximal chemical potential imbalance
\begin{equation}
\label{eq:maximal_chemical_potential_imbalance}
    \Delta\mu^\star_0(v) \coloneqq \lim_{M\rightarrow\infty} \Delta\mu_0(v) = - \chi_s \Delta s(v) - \chi_p \Delta p(v) \, .
\end{equation}
As discussed in Sec.~\ref{sec:chemical_potential_imbalance_phase_portrait_analysis}, steady states of the condensate dynamics are defined by the roots of Eq.~\eqref{eq:chemical_potential_imbalance_1D_rewritten}.
Moreover, the thermodynamic housekeeping analysis in Sec.~\ref{sec:chemical_potential_imbalance_phase_portrait_analysis} showed that stable steady states are characterized by a negative slope ${\partial_v \Delta\mu_0(v^\star) < 0}$ whereas unstable states have a positive slope ${\partial_v \Delta\mu_0(v^\star) > 0}$.
Geometrically, this means that---in stable steady states---the curve $\Delta\mu^\star_0(v)$ intersects the line ${(2 R \Delta c)/(M c_+) \, v}$ from above.
Finally, as discussed in the previous section, the largest root of Eq.~\eqref{eq:chemical_potential_imbalance_1D_rewritten} must correspond to a stable steady state.
Therefore, one can map out the bifurcation diagram simply by determining the roots.

The above criteria to observe condensate self-propulsion can only be fulfilled if the maximal chemical potential imbalance $\Delta\mu^\star_0(v)$ [Eq.~\eqref{eq:maximal_chemical_potential_imbalance}] has a band of positive (negative) values for positive (negative) velocities.
In that case, as mentioned at the beginning of this section, Eq.~\eqref{eq:chemical_potential_imbalance_1D_rewritten} is guaranteed to have non-trivial roots for ${v \neq 0}$ if the enzyme mobility is sufficiently high,
\begin{equation} 
    M > \frac{2 R \Delta c}{c_+} \, \min_{v} \frac{v}{\Delta\mu^\star_0(v)} > 0
    \, .
\end{equation}
In the following, we exploit these simple rules to improve our understanding of the onset of condensate self-propulsion.
Because the system does not have a preferred spatial direction, without loss of generality, we focus on positive droplet velocities ${v \geq 0}$.

\subsubsection{Weakly reciprocal droplets}
\label{sec:weak_reciprocity}
First, we consider a scenario where the reciprocity parameter $\Lambda$ is small.
To that end, we expand the maximal chemical potential imbalance to first order in $\Lambda \chi_{s,p} \Delta c / D$,
\begin{equation}
\label{eq:potential_imbalance_expansion}
    \Delta\mu^\star_0(v) \approx \Delta\mu^\star_0(v) \bigr\rvert_{\Lambda = 0} + \partial_\Lambda \Delta\mu^\star_0(v) \bigr\rvert_{\Lambda = 0} \, \Lambda \, ,
\end{equation}
which implicitly assumes weak enzyme-substrate and enzyme-product interactions.
Based on the discussion in the previous section, droplet self-propulsion is only possible if the maximal chemical potential imbalance $\Delta\mu^\star_0(v)$ has a band of positive values for ${v > 0}$.
To gain a clearer understanding of when this condition holds, we will next discuss the characteristic features of the two leading-order contributions to the series expansion in Eq.~\eqref{eq:potential_imbalance_expansion}.

\begin{figure}[t]
\centering
\includegraphics{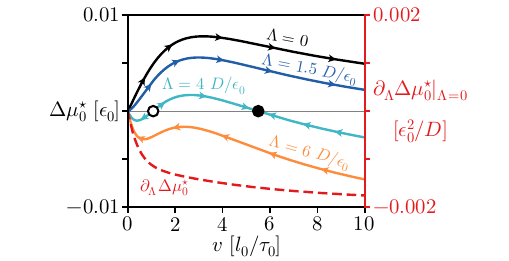}
\caption{%
Maximal chemical potential imbalance $\Delta\mu_0^\star(v)$ [Eq.~\eqref{eq:maximal_chemical_potential_imbalance}] as a function of the droplet speed $v$, for different values of the parameter $\Lambda$. 
In the limiting case ${M \to \infty}$, steady states are defined by the roots of the maximal chemical potential imbalance.
Arrows indicate how the chemical potential imbalance affects the droplet speed $v$.
Stable (filled circles) and unstable (open circles) steady states are indicated for ${\Lambda = 4 D / \epsilon_0}$.
Color code of curves is the same as in Fig.~\ref{fig:self_propulsion_reciprocal} and in Fig.~\ref{fig:velocities_lambda}.
The dashed red line shows how the maximal chemical potential imbalance is affected by the parameter $\Lambda$.
Parameters: ${\chi_s = -0.05 \, r}$, ${\chi_p = -0.01 \, r}$, ${c_- = 0.1 \, c_+}$, ${R = l_0}$, ${k_1 = k_2/c_+}$, and ${\langle s+p \rangle = c_+}$.}
\label{fig:reciprocal_expansion}
\end{figure}

\emph{Leading order: nonreciprocal limit.}
The first term in Eq.~\eqref{eq:potential_imbalance_expansion}, ${\Delta\mu^\star_0(v) \rvert_{\Lambda = 0}}$, recapitulates the limit of nonreciprocal droplets [black curve in Fig.~\ref{fig:reciprocal_expansion}], which we have discussed in our previous work~\cite{Demarchi2023}.
In this limit, an example of the substrate and product concentration profiles, ${s(z)\rvert_{\Lambda = 0}}$ and ${p(z)\rvert_{\Lambda = 0}}$, is depicted in the left panel of Fig.~\ref{fig:thermodynamic_vs_nonreciprocal}.
As these concentration profiles show, reactions and diffusion lead to depletion of substrates and enrichment of products at the trailing edge of the moving condensate, in comparison to its leading edge. 
Thus, considering Eq.~\eqref{eq:maximal_chemical_potential_imbalance}, ${\Delta\mu^\star_0(v) \rvert_{\Lambda = 0}}$ must be strictly positive for ${v > 0}$ when enzyme-substrate attraction dominates over enzyme-product interactions.
As we discussed in the previous section, the existence of a band of positive values ${\Delta\mu^\star_0(v) \rvert_{\Lambda = 0} > 0}$ for ${v > 0}$ fulfills the necessary condition for droplet motion.
Moreover, the positive slope of the curve ${\Delta\mu^\star_0(v) \rvert_{\Lambda = 0} > 0}$ at the fixed point ${v = 0}$ is indicative of an unstable state.
Hence, we conclude that the leading-order term in the series expansion of Eq.~\eqref{eq:potential_imbalance_expansion} promotes droplet self-propulsion.

It is important to note that this leading-order term is bound from above for positive droplet velocities.
This can be deduced from the fact that ${\Delta\mu^\star_0(v) \rvert_{\Lambda = 0}}$ must vanish in the following two limiting cases.
For ${v=0}$, the chemical potential imbalance vanishes because the system is symmetric under parity transformations.
In the limit ${v\to\infty}$, the substrate and product concentration profiles remain flat, as the time required to traverse one droplet diameter, $2R / v$, is much shorter than the timescales of reactions, $\tau_0$, and diffusion, $R^2 / D$, so that the densities equilibrate and become uniform.
Since the leading-order term ${\Delta\mu^\star_0(v) \rvert_{\Lambda = 0}}$ is strictly positive for positive velocities ${v>0}$, it follows that it must have a maximum for some finite value ${v > 0}$ [black curve in Fig.~\ref{fig:reciprocal_expansion}].
As a consequence, the correction term in the series expansion of Eq.~\eqref{eq:potential_imbalance_expansion} can exceed this zeroth-order contribution.

\emph{Reciprocal correction.}
The second term in Eq.~\eqref{eq:potential_imbalance_expansion}, ${\partial_\Lambda \Delta\mu^\star_0(v) \rvert_{\Lambda = 0}}$, is a correction that accounts for weak reciprocity [dashed red curve in Fig.~\ref{fig:reciprocal_expansion}].
The corresponding changes in the substrate and product%
\footnote{Note that in the equilibrium scenario depicted in Fig.~\ref{fig:thermodynamic_vs_nonreciprocal}, only substrates are present in the system.}
concentration profiles can be understood from the right panel of Fig.~\ref{fig:thermodynamic_vs_nonreciprocal}: 
Their attractive interactions with enzymes, ${\chi_{s,p} < 0}$, will favor influx at the leading edge but hinder outflux of substrates and products at the trailing edge of the droplet.
Hence, substrates and products will accumulate at the trailing edge of the condensate which acts as a moving potential barrier%
\footnote{Repulsive attractions would instead cause accumulation at the leading edge.}.
These concentration profiles, considering Eq.~\eqref{eq:maximal_chemical_potential_imbalance}, lead to ${\partial_\Lambda \Delta\mu^\star_0(v) \rvert_{\Lambda = 0}}$ being strictly negative and decreasing monotonically with increasing droplet speed, as more substrates and products accumulate at the trailing edge of the condensate.
Therefore, with increasing $\Lambda$, the correction term can cause the maximal chemical potential imbalance [Eq.~\eqref{eq:potential_imbalance_expansion}] to become strictly negative for ${v>0}$, leaving only a single stable fixed point ${v = 0}$ [solid red curve in Fig.~\ref{fig:reciprocal_expansion}].
In that case, droplet self-propulsion is completely suppressed.

\begin{figure}[t]
\centering
\includegraphics{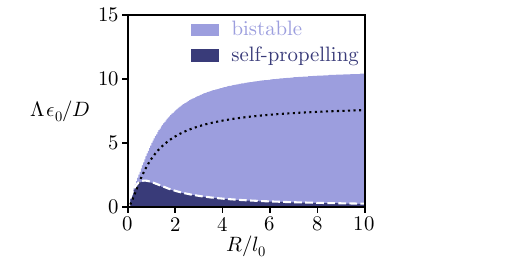}
\caption{%
Phase diagram illustrating how droplet self-propulsion depends on the condensate radius $R$ and on the mobility of substrates and products in response to a gradient in enzymes, ${\Lambda(s) =\Lambda s}$ and ${\Lambda(p) = \Lambda p}$, in the sharp interface approximation.
The light blue region shows the parameter regime where droplet self-propulsion is bistable [Eq.~\eqref{eq:condition_positive_chemical_potential_imbalance}].
The corresponding phase boundary can be crudely approximated by Eq.~\eqref{eq:condition_positive_chemical_potential_imbalance_approximation}, as shown by the black dotted line.
The dark blue region shows the spinodal parameter regime where only self-propelling condensate states are stable [Eq.~\eqref{eq:nonreciprocal_slope_criterion}].
For simplicity, we consider the limit $M\to\infty$.
Parameters: ${\chi_s = -0.05 \, r}$, ${\chi_p = -0.01 \, r}$, ${c_- = 0.1 \, c_+}$, ${k_1 = k_2/c_+}$, and ${\langle s+p \rangle = c_+}$.
}
\label{fig:phase_diagram_reciprocal}
\end{figure}

\emph{Bistability and criticality.}
In section~\ref{sec:discussion_and_test_reciprocal_theory}, we discussed that for ${\Lambda > 0}$, self-propulsion typically sets in through a subcritical bifurcation when varying the enzyme mobility $M$, and that there is a bistable region.
In the following, we use our framework to further elucidate these phenomena.
Recall that the condition for a stable steady state is that the curve $\Delta\mu^\star_0(v)$ cuts the line ${(2 R \Delta c)/(M c_+) \, v}$ from above.
To observe a supercritical bifurcation when varying the enzyme mobility $M$, the fixed point ${v = 0}$ must lose its stability concomitantly with the emergence of a new branch of stable fixed points for ${v \geq 0}$.
An important feature of a supercritical transition is that the velocity does not suddenly jump to a finite value when varying the enzyme mobility $M$ for example, but instead continuously increases from ${v=0}$.
Geometrically, this is only possible if the maximal chemical potential imbalance $\Delta\mu^\star_0(v)$ grows but curves downward, ${\partial_v^2 \Delta\mu^\star_0(v)\rvert_{v=0} \leq 0}$, when the velocity is increased starting from the fixed point ${v=0}$.
Using Mathematica, we found that the leading order term in the series expansion in Eq.~\eqref{eq:potential_imbalance_expansion} always has vanishing curvature, ${\partial_v^2 \Delta\mu^\star_0(v)\rvert_{\Lambda=0, v=0} = 0}$.
While we were not able to completely map out the correction term, we found that it curves upward as a function of the droplet velocity, for small droplet radii $R$ and for the parameters studied in Fig.~\ref{fig:phase_diagram_reciprocal}. 
This further supports the conclusion that for ${\Lambda > 0}$, in a wide range of parameters, the onset of condensate self-propulsion is subcritical.

For sufficiently high enzyme mobility $M$, a subcritical bifurcation occurs when the fixed point ${v=0}$ becomes unstable.
This is only possible if the maximal chemical potential imbalance $\Delta\mu^\star_0(v)$ has a positive slope at the fixed point ${v=0}$.
Using the series expansion in Eq.~\eqref{eq:potential_imbalance_expansion}, this leads to  
\begin{equation}
\label{eq:nonreciprocal_slope_criterion}
    \Lambda < 
    - \frac{\partial_v \Delta\mu^\star_0(v) \bigr\rvert_{\Lambda = 0, v=0}}
    {\partial_v \partial_\Lambda \Delta\mu^\star_0(v) \bigr\rvert_{\Lambda = 0, v=0}} \, ,
\end{equation}
where we have used the fact that ${\partial_\Lambda \Delta\mu^\star_0(v)} \leq 0$.
In Fig.~\ref{fig:phase_diagram_reciprocal}, the dark blue region shows when this criterion is fulfilled for different values of the reciprocity parameter $\Lambda$ and the condensate radius $R$.
Even if the slope criterion Eq.~\eqref{eq:nonreciprocal_slope_criterion} is not fulfilled, the system can still be bistable as we explain next. More specifically, to permit condensate self-propulsion for sufficiently high enzyme mobility, it suffices if the maximal chemical potential imbalance $\Delta\mu^\star_0(v)$ has a positive maximum [cf. light blue curve in Fig.~\ref{fig:reciprocal_expansion}],
\begin{equation}
\label{eq:condition_positive_chemical_potential_imbalance}
    \max_{v>0} \Delta\mu^\star_0(v) > 0 \, ,
\end{equation}
for positive velocities.
In Fig.~\ref{fig:phase_diagram_reciprocal}, the light blue region shows when  criterion Eq.~\eqref{eq:condition_positive_chemical_potential_imbalance} is fulfilled for different values of the reciprocity parameter $\Lambda$ and the condensate radius $R$.
The inequality~\eqref{eq:condition_positive_chemical_potential_imbalance} is automatically satisfied when the maximum of the leading-order term in Eq.~\eqref{eq:potential_imbalance_expansion} is larger than the saturation value of the correction, which leads to the weaker constraint
\begin{equation}
\label{eq:condition_positive_chemical_potential_imbalance_approximation}
    \Lambda \lesssim -\frac{\max_v \Delta\mu^\star_0(v)\bigr\rvert_{\Lambda = 0}}{\lim_{v\rightarrow\infty} \partial_\Lambda \Delta\mu^\star_0(v)\bigr\rvert_{\Lambda = 0}} \, .
\end{equation}
Here, we have used the fact that ${\partial_\Lambda \Delta\mu^\star_0(v) \leq 0}$.
This approximation gives a lower bound for the critical value of $\Lambda$ below which self-propulsion is possible [black dotted line in Fig.~\ref{fig:phase_diagram_reciprocal}].

In summary, we have explored how the magnitude of the reciprocal interactions, quantified by $\Lambda$, controls the onset of condensate self-propulsion.
In the bistable region, it requires a sufficiently large perturbation of the system to excite the condensates into a self-propelling state.
Importantly, all of our results so far suggest that the reciprocity parameter $\Lambda$, and hence enzyme-substrate and enzyme-product interactions, must be weak to observe condensate motion.
How can we relax this constraint?

\subsubsection{Droplets in the strongly reciprocal regime}
\label{sec:strong_reciprocity}
So far, we have shown that droplet self-propulsion can be inhibited if the reciprocity parameter $\Lambda$ is sufficiently large.
Then, is it possible at all to observe droplet self-propulsion when ${\Lambda \, \epsilon_0 / D \rightarrow \infty}$? 
This would correspond to a limit at which enzyme-substrate and enzyme-product interactions are strong, so that the effects due to reciprocity become very important.
To understand the particular significance of this case, we again use the Einstein-Smoluchowsky relation ${D=\Lambda \, k_B T}$ to relate the mobility of substrates and products to their diffusion coefficient in the case of thermal Brownian motion~\cite{Frey2005}.
The horizontal axis in Fig.~\ref{fig:self_propulsion_reciprocal} can then be interpreted as ${M \epsilon_0 / D = (M / \Lambda) \, \epsilon_0 / (k_B T)}$ and the vertical axis as ${\Lambda \epsilon_0 / D = \epsilon_0 / (k_B T)}$.
Thus, for the parameters discussed so far, weak enzyme-substrate and enzyme-product interactions, with energy scale ${\epsilon_0 \sim k_B T}$, would necessitate a large ratio of mobilities ${(M / \Lambda)}$ to induce droplet self-propulsion.
This can be the case, for example, if the enzymes correspond to small proteins (such as transcription factors and components of the transcription machinery) and the products correspond to large RNA molecules%
\footnote{To treat this scenario, one would need to relax the assumption of substrates and products having equal diffusion coefficients. More specifically, the produced RNA polymer would have much smaller diffusion coefficient than the substrate nucleotides. In addition to this leading-order effect, at sufficiently high concentrations, the dynamics of substrates and products will cease to resemble the behavior of an ideal solution.
In that case, one will also need to take into account the different polymerization factors by using Flory-Huggins theory.}%
\textsuperscript{,}%
\footnote{If the enzymes are proteins that cleave RNA polymers, then the products would have higher diffusion coefficient than the substrates.}.

To explore if droplet self-propulsion is possible for a small ratio of mobilities ${(M / \Lambda)}$, we will now study the extreme limit ${\Lambda \, \epsilon_0 / D \rightarrow \infty}$.
So far, we have focused on a scenario where both substrates and products show attractive interactions with enzymes, with ${\chi_s < \chi_p < 0}$.
As shown in Appendix~\ref{appendix:large_lambda}, and in agreement with our results in Sec.~\ref{sec:discussion_and_test_reciprocal_theory} and Sec.~\ref{sec:weak_reciprocity}, in this case the state ${v=0}$ will always remain stable in the limit ${\Lambda \, \epsilon_0 / D \rightarrow \infty}$.
Hence, we will now modify the above assumptions, keeping the enzyme-substrate interactions attractive, ${\chi_s < 0}$, but now assuming that enzyme-product interactions are repulsive, ${\chi_p > 0}$; 
both are assumed to be weaker than enzyme-enzyme interactions so that the sharp interface approximation remains valid.
Moreover, we consider condensates where the concentration of enzymes is enriched inside the droplet, ${\Delta c > 0}$.
With these assumptions, one can determine the concentration jumps for substrates and products at the droplet interface, Eq.~\eqref{eq:concentration_jump_main}, in the limit ${\Lambda \, \epsilon_0 / D \rightarrow \infty}$:
\begin{subequations}
\label{eq:concentration_jump_reciprocal_limit}
\begin{align}
    \frac{s\vert_\text{in}}{s\vert_\text{out}} 
    &= 
    \exp \biggl[ -\frac{\Lambda \, \chi_s \, \Delta c}{D} \biggr] 
    \to \infty
    \, , 
    \\
    \frac{p\vert_\text{in}}{p\vert_\text{out}} 
    &= 
    \exp \biggl[ -\frac{\Lambda \, \chi_p \, \Delta c}{D} \biggr] 
    \to 0
    \, .
\end{align}
\end{subequations}
This limit implies ${p\vert_\text{in} = 0}$ and ${s\vert_\text{out} = 0}$, which considerably simplifies the further analysis.
Next, similar as in Sec.~\ref{sec:weak_reciprocity}, we study the stability of the fixed point at ${v = 0}$. 
A non-moving droplet state, ${v = 0}$, is unstable when the slope of the chemical potential imbalance is positive, 
\begin{equation}
\label{eq:criterion_self_propulsion_strong}
    \partial_v \Delta\mu_0(v) \rvert_{v=0} = 
    \partial_v \Delta\mu_0^\star(v) \rvert_{v=0} 
    - \frac{2R \Delta c}{M c_+} > 0 \, ,
\end{equation}
where the first equality follows from Eq.~\eqref{eq:chemical_potential_imbalance_1D_rewritten}.
We solve for the substrate and product concentration profiles, as discussed in Appendix~\ref{sec:substrate_interaction_profile_reciprocal} but now with the simplified boundary conditions Eq.~\eqref{eq:concentration_jump_reciprocal_limit}.
After substituting these profiles in the maximal chemical potential imbalance Eq.~\eqref{eq:maximal_chemical_potential_imbalance}, one has\footnote{We carried out the analytic calculations with the computer algebra system Mathematica~\cite{Mathematica} and provide the corresponding notebooks in Ref.~\cite{CodeGithub}.}
%
\begin{multline}
\label{eq:maximal_chemical_potential_imbalance_strong_reciprocity}
    \partial_v\Delta\mu_0^\star(v)\rvert_{v=0} = 
    \frac{2 R \, n_\infty \chi_s}{D ( l_0^2 - l_+^2 )} \color{gray}\times\color{black} \\ 
    \color{gray}\times\color{black} \left[
    l_- l_+ \coth( R/l_+ ) + l_+^2 - l_0^2
    \right] \, ,
\end{multline}
where ${l_\pm = \sqrt{D/(k_1 c_\pm + k_2)}}$ are the diffusion lengths inside and outside the condensate, respectively.
For inequality~\eqref{eq:criterion_self_propulsion_strong} to be fulfilled, the maximal chemical potential imbalance must have a positive slope [Eq.~\eqref{eq:maximal_chemical_potential_imbalance_strong_reciprocity}].
Note that, since ${l_\pm < l_0 = \sqrt{D/k_2}}$ and $\chi_s < 0$, the factor in front of the square brackets in  Eq.~\eqref{eq:maximal_chemical_potential_imbalance_strong_reciprocity} is always negative.
Therefore, a first necessary criterion for the emergence of droplet self-propulsion follows by requiring that the expression in the square brackets in Eq.~\eqref{eq:maximal_chemical_potential_imbalance_strong_reciprocity} is also negative, so that the maximal chemical potential imbalance has a positive slope.
This suggests that droplets can exhibit self-propulsion if the droplet size $R$ is sufficiently large. 
Because ${\coth(R/l_+) > 1}$ and ${\chi_s < 0}$, Eq.~\eqref{eq:maximal_chemical_potential_imbalance_strong_reciprocity} has the following bound:
\begin{equation}
    \partial_v\Delta\mu_0^\star(v)\rvert_{v=0} \leq 
    \frac{2 R \, n_\infty \chi_s}{D ( l_0^2 - l_+^2 )}  
    \left[
    l_- l_+ + l_+^2 - l_0^2
    \right] ,
\end{equation}
leading to the additional criterion ${l_0^2 > l_- l_+ + l_+^2}$.
This condition can be rewritten by substituting the expressions for the diffusion lengths ${l_\pm = \sqrt{D/(k_1 c_\pm + k_2)}}$, leading to 
\begin{equation}
\label{eq:reciprocal_strong_inequality}
    \left(\frac{k_1 c_+}{k_2}\right)^2 > \frac{1+k_1 c_+ / k_2}{1+k_1 c_- / k_2} \geq 1 + \frac{k_1 c_+}{k_2} \, ,
\end{equation}
where the last inequality approximately becomes an equality in the limit of strong phase separation, ${c_-/c_+ \approx 0}$.
In this limit, one can solve the quadratic inequality~\eqref{eq:reciprocal_strong_inequality} exactly, leading to the criterion ${k_1 c_+ / k_2 > (\sqrt{5} + 1)/2}$ for the reaction rates.
Thus, the turnover rate of substrates must be sufficiently fast for droplets to show self-propulsion.

Finally, substituting Eq.~\eqref{eq:maximal_chemical_potential_imbalance_strong_reciprocity} into the inequality~\eqref{eq:criterion_self_propulsion_strong} and solving for the enzyme mobility leads to a lower enzyme mobility bound,
\begin{equation}
    \frac{M}{\Lambda} \geq -\frac{k_B T}{\chi_s n_\infty} \frac{\Delta c}{c_+} \left[1 - \frac{ l_- l_+ \coth(R/l_+) }{l_0^2 - l_+^2} \right]^{-1} \, ,
\end{equation}
above which condensates will begin to self-propel; note that we have again used the Einstein-Smoluchowsky relation ${D = \Lambda \, k_B T}$.
Thus, the critical ratio between the mobility of enzymes and the mobility of substrates and products can be lowered by decreasing the enzyme's propensity to phase separate, $\Delta c$, or by increasing the interaction strength, $\chi_s$.
Importantly, the critical mobility ratio ${M / \Lambda \propto -k_B T / (\chi_s n_\infty)}$ can become very small in the limit of strong enzyme-substrate interactions ${\chi_s \ll -k_B T / n_\infty}$. 
To summarize this section, we have found that, in general, droplet self-propulsion requires sufficiently fast reaction rates, sufficiently large condensates, and sufficiently large enzyme mobility.

\section{Discussion}
\label{sec:discussion}

We have studied the non-equilibrium dynamics of biomolecular condensates (droplets) containing a high concentration of enzymes.
Through their catalytic role in biochemical reactions, these enzymes establish a spatial framework that governs the arrangement of substrates and products.
In turn, the resulting substrate and product concentration profiles drive the enzyme flows through molecular interactions.
This interplay between non-equilibrium reactions and reciprocal biomolecular interactions can give rise to novel phenomena, such as droplet self-propulsion, which we have here studied in depth.

We have identified the criteria for droplet self-propulsion in the two opposing limits of weak or strong enzyme-substrate and enzyme-product interactions.
These limits correspond to small or large values of the reciprocity parameter $\Lambda$, which represents the magnitude of the Onsager coefficient (mobility) defining the extent of the currents induced by gradients in the chemical potentials of substrates and products.
In the limit of weak interactions, our first analysis indicated that droplet self-propulsion requires a very small ratio of mobilities ${\Lambda / M \ll 1}$ [Fig.~\ref{fig:self_propulsion_reciprocal}], where $M$ is the mobility of enzymes.
This could, for example, apply to experiments where self-propelling droplets leave a long low-pH trail by producing acid molecules~\cite{Hanczyc2011}.
However, there is also experimental evidence for the collective migration of urease-containing droplets~\cite{Mirco2023, JambonPuillet2023} which seem to disobey the above restriction. 
Motivated by these experiments, we used our theory to investigate a scenario where the reciprocity parameter $\Lambda$ is large or, in other words, enzyme-substrate and enzyme-product interactions are strong. 
Our analysis revealed that, in this case droplet self-propulsion is possible for repulsive enzyme-product interactions, $\chi_p > 0$, if the reaction rates are sufficiently fast and the condensate is sufficiently large.
These results qualitatively agree with Ref.~\cite{JambonPuillet2023}, where motion was only observed for droplets whose radius exceeded \SI{20}{\micro\meter}.

Interestingly, we have found that liquid-like droplets generally move faster than solid-like condensates. 
The reason for this finding is that droplets can translocate simply by increasing solubility at one interface and decreasing solubility at the opposing interface~\cite{Weber2017}.
Using differences in solubility to dissolve one side (inward motion) and grow the opposing side (outward motion) of the condensate will cause droplet motion without a net transport of mass.
In contrast, the motion of a solid-like condensate implies a transport of its entire mass, and will hence be limited by the mobility of its molecular components.

\emph{Adiabatic elimination scheme.}
To gain these insights, we have developed a self-consistent sharp-interface theory that describes the dynamics of active phase-separated systems.
Using an implicit adiabatic elimination scheme---based on the assumption that the concentration profile of the enzymes is always maintained adiabatically in a steady state---we inferred the chemical potential and currents of the enzymes necessary to maintain a state with a given droplet shape and velocity.
Finally, we determined which value of the droplet speed is thermodynamically consistent, by calculating the rate of power dissipation. 
The power dissipation reveals if a given droplet state can be realized without supplying additional energy or removing excess energy from the system.
This thermodynamic housekeeping analysis suggests that increasing the droplet speed $|v|$ compared to a stable steady state would require energy influx, whereas decreasing the droplet speed $|v|$ would require energy outflux.

\emph{Droplet shape.}
We restricted our analysis to a scenario where the condensates remain in a round shape.
This is generally a good approximation near the onset of rotational symmetry breaking and self-propulsion, or if the surface tension of the condensates is large; the latter case is realized for strong enzyme-enzyme interactions.
However, the approximation of a round droplet shape will need to be relaxed if the enzyme-enzyme interactions are sufficiently weak compared to the enzyme-substrate and enzyme-product interactions.
To then ensure that the droplet surface tension remains a finite-valued material parameter, the interface stiffness of the droplet will need to be inversely proportional to the strength of enzyme-enzyme interactions.
Then, droplets could deform into non-spherical shapes with non-uniform interface curvature, thereby leading to Laplace pressure variations that affect the enzyme currents.
Such a generalization would require the derivation of conditions for the droplet shape, which will be addressed in future studies.
This would aid future studies in investigating changes and instabilities in droplet shape~\cite{Zwicker2016}.

\emph{Analogy to fluid mechanics.}
At the heart of our theoretical analysis, we studied enzyme droplets in a moving steady state by taking a sharp interface limit.
To ensure stationarity of the concentration profiles in the co-moving frame, we constructed a chemical potential such that the enzyme currents were forced to be divergence-free.
This closely resembles the strategy used for incompressible fluids, where an implicitly defined finite pressure field enforces the fluid incompressibility condition.
To make this analogy more explicit, we will now compare the overdamped dynamics of the enzyme currents [shown as stream plot in Fig.~\ref{fig:self_consistency_example}(a)] to the inviscid hydrodynamics described by Euler's equations~\cite{Book:Landau_FluidMech}. 
For an inviscid fluid with zero shear viscosity, the fluid velocity $\mathbf{u}$ is determined by a balance between inertial forces and local driving by applying an external force $\mathbf{g}$, with a pressure field $p$ enforcing fluid incompressibility:
\begin{subequations}    
\begin{align}
    \partial_t (\rho \, \mathbf{u}) 
    &= 
    -\boldsymbol\nabla p + \mathbf{g} \, , 
    \\
    \boldsymbol\nabla \cdot \mathbf{u} 
    &= 
    0 \, ,
\end{align}
\end{subequations}
where $\rho$ is the fluid density and fluid convection is neglected. 
In analogy, on the right-hand side of Eq.~\eqref{eq:effective_euler-a}, the chemical potential $\mu_0$ plays the role of an effective pressure field that enforces incompressibility as described by Eq.~\eqref{eq:effective_euler-b}, while the remaining terms are analogous to an applied force.
In contrast to Euler's equations, however, the current on the left-hand side of Eq.~\eqref{eq:effective_euler-a}, when divided by the enzyme mobility, corresponds to drag friction forces instead of inertial forces, and is therefore irreversible.

Finally, we note that the description of our model in terms of currents can naturally incorporate hydrodynamic interactions.
We expect that our self-consistent sharp interface theory can be readily applied to such a scenario.
By choosing the mobility function $M(c)$ to model a solid where only the dense phase is mobile, and the force field $\mathbf{f}$ to model interactions with a neutral or charged solute, for example, one could compare the diffusiophoresis~\cite{Review::MarbachBocquet2019} of solid colloids to that of liquid droplets.
Even in the absence of hydrodynamic interactions, our results already suggest that liquid-like condensates can be transported much faster than solid particles, because such a transport only requires moving the concentration profile instead of every molecule contained in the condensate.

\emph{Reciprocal interactions.}
Using the theoretical framework developed here, we have re-examined the phenomenon of self-propulsion of droplets, which was studied in our previous work~\cite{Demarchi2023} in the non-reciprocal limit (${\Lambda = 0}$), for cases where reciprocity is restored.
We found that for ${\Lambda>0}$ interactions with enzymes generally lead to discontinuities in the substrate and product concentration profiles.
As reciprocity moves the system closer to thermodynamic equilibrium, additional conditions must be met to observe the self-propulsion of the droplets:
\begin{enumerate*}[label=(\roman*)]
\item For weak enzyme-substrate and enzyme-product interactions, the reciprocity parameter must lie below a certain threshold, Eq.~\eqref{eq:condition_positive_chemical_potential_imbalance_approximation}, to observe self-propulsion.
In general, the onset of self-propulsion will then be discontinuous (supercritical) as one varies, for example, the enzyme mobility $M$. 
A second, lower threshold determines if the system is bistable (binodal regime), or if the onset of self-propulsion is spontaneous (spinodal regime).
The onset of droplet self-propulsion only becomes continuous (supercritical) in the nonreciprocal limit (${\Lambda = 0}$).
\item In principle, droplet self-propulsion is also possible for large reciprocity parameters, ${\Lambda \to \infty}$, if enzyme-product interactions are repulsive, ${\chi_p > 0}$.
However, the size of the droplet must then be considerable, and the non-equilibrium reaction rates must be sufficiently fast.
\end{enumerate*}
Similar to the non-reciprocal limit in Ref.~\cite{Demarchi2023}, as briefly discussed in Appendix~\ref{appendix:positioning_coexistence}, in the reciprocal case one can also observe droplet positioning, elongation, and division.

\emph{Fisher waves and other model systems.}
The theory of moving droplet interfaces developed here was in part inspired by the analysis of Fisher waves~\cite{FISHER1937}, which model the expansion of growing populations.
However, in contrast to the model studied in the present work, which was derived based on thermodynamic arguments and incorporated phase separation \emph{a priori}, population dynamics are arguably far from thermal equilibrium and described by phenomenological models.
It would therefore be interesting to test if the ideas underlying the framework presented in this work can be adapted to front propagation in such far-from-equilibrium systems.
If that is the case, then the derivation of balance conditions for the arising pseudo-chemical potentials could pave the way to borrow powerful tools from non-equilibrium thermodynamics for gaining new insights into ecological systems.
However, we also note that a central aspect of our analysis was the conservation law for the total mass of enzymes, while population dynamics models lack such conservation laws.
Therefore, we expect that the present framework can be more readily adopted for the analysis of non-reciprocal mass-conserved models~\cite{Saha2020, Brauns2023} or mass-conserved reaction-diffusion systems with advection~\cite{Wigbers2020}.

\section{Conclusion and outlook}
The interplay of reciprocal biomolecular interactions that cause phase separation, coupled with out-of-equilibrium chemical reactions, is a widespread organizational motif in living cells. 
By developing a versatile sharp interface theory we have gained insight into the specific conditions under which this motif can lead to the self-propulsion of condensates.
The mechanism underlying condensate motion can be understood as a gradient in solubility which leads to translocation of droplets~\cite{JambonPuillet2023}. 
More specifically, attractive interactions with substrates in the surrounding solution and repulsion with products in the condensate, locally increase the solubility of the enzymes (that is, how much the enzymes are driven towards the surrounding solution).
This leads to the dissolution of the trailing edge of the droplet, which is enriched in products and depleted in substrates, and growth of the leading edge of the droplet.
In addition, the net gradient of substrates across the entire droplet attracts enzymes towards the leading edge of the condensate, where the substrate concentration is higher.

The ability of biomolecular condensates to migrate along concentration gradients, which was recently recognized theoretically~\cite{Weber2017, Demarchi2023, Hafner2023} and demonstrated experimentally for pH~\cite{Hanczyc2011, Mirco2023, JambonPuillet2023} or salt~\cite{Doan2023} gradients, is reminiscent of the diffusiophoresis of colloids~\cite{Review::MarbachBocquet2019}.
For colloidal diffusiophoresis, however, hydrodynamic shear stresses play an essential role because interactions between the colloidal surface and the inhomogeneously distributed solute lead to an effective surface tension gradient~\cite{Review::MarbachBocquet2019}.
While we have here neglected fluid viscosity, its consideration in future studies would make it possible to further elucidate the parallels and differences between droplet and colloid motion.
Moreover, accounting for fluid mechanics would provide a more accurate description of experimental systems in which condensates move due to chemical gradients such as salt or pH variations.

The response to an applied gradient in concentration, or mechanical (that is, viscoelastic) properties of the surrounding medium is a unifying theme among soft and living matter.
In addition to chemical gradients which drive diffusiophoresis and cellular chemotaxis, for example, it was shown that droplets~\cite{Style2013, Rosowski2020} and cells~\cite{Sunyer2020} can also migrate along stiffness gradients.
Given the success of phase field models in describing cell migration~\cite{Shao2010, Shao2012, Ziebert2012, Ziebert2013, Camley2013, Lober2014}, we hypothesize that some of the ideas developed in the present manuscript could also apply to cellular dynamics.

\begin{acknowledgments}
We thank Dominik Schumacher and Lotte S\o{}gaard-Andersen for helpful discussions. 
We acknowledge financial support by the German Research Foundation (DFG) through TRR 174 (Project ID No.~269423233) and SFB1032 (Project ID No.~201269156) and the Excellence Cluster ORIGINS under Germany’s Excellence Strategy (EXC-2094-390783311). 
During his time at the Ludwig-Maximilians-Universit\"at M\"unchen, AG was supported by a DFG fellowship through the Graduate School of Quantitative Biosciences Munich (QBM). 
During his time at the Massachusetts Institute of Technology, AG was supported by the National Science Foundation (NSF) through grant number 2044895 and by an EMBO Postdoctoral Fellowship (ALTF 259-2022).
During his time at Sorbonne University, LD has received funding from the European Union’s Horizon 2020 research and innovation programme under the Marie Sk\l{}odowska-Curie grant agreement No. 860949.
IM has received funding from the European Union's Framework Programme for Research and Innovation Horizon 2020 under the Marie Sk\l{}odowska-Curie Grant Agreement No. 754388 (LMU Research Fellows) and from LMU excellent, funded by the Federal Ministry of Education and Research (BMBF) and the Free State of Bavaria under the Excellence Strategy of the German Federal Government and the L\"ander.
EF was supported by the European Union (ERC, CellGeom, project number 101097810) and the Chan-Zuckerberg Initiative (CZI).
\end{acknowledgments}

\bibliography{droplets}

\clearpage\newpage

\appendix

\section{Effect of interactions on reaction rates}
\label{appendix:transition_state}

The rate constant $k_2$ of the reaction path 
\begin{equation}
    \ce{$P$ + $F$ ->[$k_2$] $S$ + $W$}
\end{equation} 
depends on the height of the corresponding potential barrier~\cite{ZwickerReview2022, Demarchi2023}.
This potential barrier depends on the energy of the transition state of the reaction and could therefore be affected by enzyme-substrate and enzyme-product interactions. 
In that case, the reaction rate would also depend on the local concentration of enzymes, even though they do not partake in this reaction.
Hence, one needs to specify how the molecules in the transition state (from product to substrate) interact with the enzymes.
We here assume that the molecules in the transition state still resemble products, so that the corresponding interaction energy,  at the maximum of the potential barrier, is the same as when product and fuel are still separated\footnote{
For instance, this scenario holds when products do not interact with enzymes and the chemical potential of the transition state remains constant~\cite{Demarchi2023}.}.
Thus, the height of the free energy barrier is set by the difference in the internal  energy of the molecules, an intensive quantity that can be taken constant, so that $k_2$ is concentration-independent. 

\section{Relative error}
\label{appendix:error}
\begin{figure}[tb]
\centering
\includegraphics{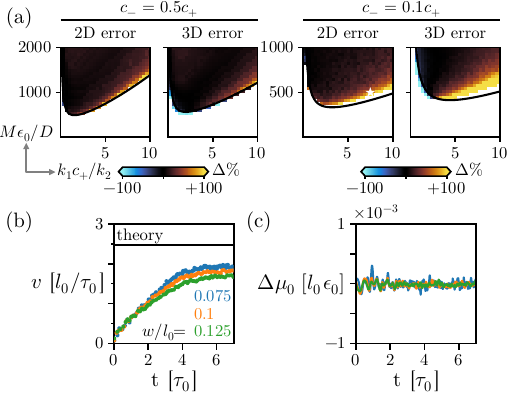}
\caption{%
\label{fig:self_consistency_test_comparison}%
\textbf{(a)} Relative error between the predictions of the sharp interface theory and the full FEM simulations shown in Fig.~\ref{fig:self_consistency_test}.
\textbf{(b)} Comparison between predicted and measured droplet speed for the parameter combination indicated by the star in Fig.~\ref{fig:self_consistency_test} and in panel (a).
Larger interface widths show larger deviation of the steady-state droplet velocity (late times) in our FEM simulations from theoretical prediction (black line).
\textbf{(c)} In all cases, the chemical potential imbalance vanishes, as posited by our theory.
}
\end{figure}

Because the droplet speed drops to zero at the phase boundary of the self-propulsion instability, the absolute error $v_\text{theory} - v_\text{simulation}$ is not necessarily a good measure of the fidelity of our predictions.
Instead, we use the relative error $v_\text{theory}/v_\text{simulation} - 1$, as shown in Fig.~\ref{fig:self_consistency_test_comparison}.
Note that the relative error can become very large for small velocities, even when the absolute error is small.
The relative error typically grows for large reaction rates because the corresponding diffusion lengths, ${l_\pm = \sqrt{D/(k_1 c_\pm + k_2)}}$, can become comparable to the droplet interface width in the simulations.
This is likely the case because the sharp interface approximation, where the interface width is taken to zero, can be invalidated when ${l_\pm \to 0}$.
In particular, in this case, the concentration profile of enzymes in the vicinity of the droplet interface will be perturbed by the local concentration gradients of substrates and products.
In agreement with these arguments, when we reduce the interface width in our FEM simulations, we observe that the droplet speed increases and thus the error of our theoretical prediction decreases.
However, note that only decreasing the interface width as control parameter (that is, decreasing the interfacial stiffness $\kappa$) will also reduce the surface tension of the condensate [Eq.~\eqref{eq:condensate_surface_tension}].
If the surface tension is very small, then we cannot assume the geometry of the droplet to remain spherical.
These limitations notwithstanding, the sharp interface theory reproduces the droplet speed with reasonable quantitative accuracy.

\section{Initial conditions and dissolution}
\label{appendix:initial_conditions}
In our sharp-interface theory, we assumed the concentration in the enzyme-rich droplets to be $c_+$ and the concentration in the enzyme-poor surroundings to be $c_-$.
In our full FEM simulations, however, there are two mechanisms that lead to a slight change in these concentration values.
The first correction arises from the Laplace pressure due to the surface tension of the droplet interface~\cite{Bray1994}, 
\begin{equation}
    \gamma = \frac{1}{6} (\Delta c)^2 r \, w \, .
\end{equation}
The second correction arises from the interactions between enzymes and substrates as well as products.
For very large droplets, and in the nonreciprocal limit ${\Lambda = 0}$, one can approximate the enzyme and substrate concentration profiles as piecewise constant, with their local reactive equilibria set by
\begin{subequations}
\begin{align}
    s^\star(c) &= \frac{n}{1 + k_1 c / k_2} \, , \\
    p^\star(c) &= n - s^\star(c)  \, .
\end{align}
\end{subequations}
The total chemical potential of enzymes is then locally given by
\begin{multline}
\label{eq:modified_chemical_potential}
    \bar{\mu} (c)
    = 
    r 
    \left[ - (c-\tilde c) + \frac{4}{\Delta c^2} (c-\tilde c)^3 
    \right] \\
    + \chi_s s^\star(c) + \chi_p p^\star(c) \, .
\end{multline}
The last two terms in Eq.~\eqref{eq:modified_chemical_potential} signify that interactions with substrates and products modify the chemical potential of enzymes.
This, in turn, affects the chemical potential balance and the osmotic pressure balance conditions at the interface of the condensate, as will be specified next.
More specifically, the enzyme concentration inside the condensate, $c_\text{in}$, and the enzyme concentration outside the condensate, $c_\text{out}$, must satisfy the following Maxwell construction~\cite{Bray1994},
\begin{subequations}
\begin{align}
    \mu(c_\text{in}) &= \mu(c_\text{out}) \, \\
    \int_{c_\text{out}}^{c_\text{in}} \!\! dc \, \mu(c) + \frac{(d-1) \gamma}{R} &= c_\text{in} \mu(c_\text{in}) - c_\text{out} \mu(c_\text{out}) \, ,
\end{align}
\end{subequations}
where the first equation represents a balance of chemical potentials and the second equation represents a balance of osmotic pressures.
These two conditions give, for large droplets, the binodal region which sets the range of average enzyme concentrations, $\frac{1}{V} \int \! d^d x \, c(\mathbf{x}) \in [c_\text{out}, c_\text{in}]$, for which a phase-separated state is thermodynamically stable.
Note that, when $\chi_{s,p}\approx 0$ and the interface tension $\gamma$ is sufficiently small or the droplet is sufficiently large, one can well approximate ${c_\text{in} \approx c_+}$ and ${c_\text{out} \approx c_-}$.

In general, attractive interactions between enzymes and substrates, which are depleted in the condensate, will drive enzyme currents towards the enzyme-poor region.
If the enzyme-substrate attraction is sufficiently strong compared to enzyme-product interactions, ${\chi_s < \chi_p}$, which is also a necessary condition for droplet self-propulsion, then one expects that the binodal region should shrink due to these interactions.
This is a conceptually very similar mechanism to the one proposed in Ref.~\cite{JambonPuillet2023}, where condensates swim towards regions where their dissolution is favored.
To ensure that the average enzyme concentration in the simulation domain lies within the binodal region, in our full 2D and 3D FEM simulations we chose the initial enzyme concentration values to be $c_\text{in}$ inside the condensate and $c_\text{out}$ outside.
If one instead chooses $c_+$ and $c_-$ for the initial enzyme concentration values, then for sufficiently large domains the average enzyme concentration will approach $c_-$ and thus leave the binodal region.
Then, droplets will dissolve after an initial period of transient motion [Supplemental Video~3].

\section{Substrate and product concentration profiles}
\label{sec:substrate_interaction_profile_reciprocal}

\subsection{Reciprocal interactions induce concentration jumps}
\label{sec:reciprocal_interactions_jump_kink}

A defining feature of taking the sharp interface limit is that the enzyme concentration profile becomes piecewise constant.
This means that, in Eq.~\eqref{eq:system_comoving} for the steady state substrate and product concentration profiles in the co-moving frame, the enzyme concentration gradient is singular at the droplet interface,
\begin{equation}
    \boldsymbol\nabla c = -\Delta c \, \delta(r-R) \, \mathbf{\hat{e}}_\perp \, ,
\end{equation}
where $\mathbf{\hat{e}}_\perp$ is the local unit normal vector.
Everywhere else in the domain, the enzyme concentration profile is approximately flat.
Hence, the terms proportional to the reciprocity parameter $\Lambda$ only play a role at the droplet interface. 
Moreoever, the reaction rates $k_1 \, c(\vec{x}) \equiv k_1 \, c_\pm$ become uniform inside and outside the droplet, respectively.
This leads to two simpler problems on the domains inside and outside the droplet, 
\begin{subequations}
\label{eq:system_comoving_subdomain}
\begin{align}
    \label{eq:substrates_comoving_subdomain}
    0 
    &= 
    \boldsymbol\nabla \cdot 
    \left( \mathbf{v} \, s + D \, \boldsymbol\nabla s  
    \right) - k_1 \, c_\pm \, s + k_2 \, p \,, \\
    \label{eq:products_comoving_subdomain}
    0 
    &= 
    \boldsymbol\nabla \cdot 
    \left( \mathbf{v} \, p + D \, \boldsymbol\nabla p  
    \right) + k_1 \, c_\pm \, s - k_2 \, p \,,
\end{align}
\end{subequations}
which, as will be discussed next, need to be connected by appropriate boundary conditions across the droplet interface.

To ensure particle number conservation of each species, the respective particle fluxes must be continuous at the droplet interface.
Akin to the domain wall theory used in studies of the totally asymmetric simple exclusion process~\cite{Kolomeisky1998, Parmeggiani2003}, this amounts to a balance between changes in diffusion and mass flux:
\begin{subequations}
\label{eq:concentration_kink}
\begin{align}
    \label{eq:concentration_kink_substrate}
    D \bigl[\boldsymbol\nabla s\vert_\text{in} - \boldsymbol\nabla s\vert_\text{out}\bigr]\cdot\mathbf{\hat{e}}_\perp &= -  \mathbf{v}\cdot\mathbf{\hat{e}}_\perp \, \bigl[ s\vert_\text{in} - s\vert_\text{out} \bigr] \, , \\
    \label{eq:concentration_kink_product}
    D \bigl[\boldsymbol\nabla p\vert_\text{in} - \boldsymbol\nabla p\vert_\text{out}\bigr]\cdot\mathbf{\hat{e}}_\perp &= - \mathbf{v}\cdot\mathbf{\hat{e}}_\perp \, \bigl[ p\vert_\text{in} - p\vert_\text{out} \bigr] \, ,
\end{align}
\end{subequations}
where $\mathbf{\hat{e}}_\perp$ denotes the unit vector normal to the droplet interface.
This relation can also be formally derived by integrating Eq.~\eqref{eq:system_comoving} over an infinitesimal line segment which crosses the sharp droplet interface.
Here $\rvert_\text{in}$ and $\rvert_\text{out}$ indicate the inner and outer side of the droplet interface, respectively.
Thus, the substrate and product concentration gradients on both sides of the droplet interface will match when the droplet is at rest (${v = 0}$) or when the concentration profiles are continuous at the interface.
However, as we show next, reciprocal interactions (${\Lambda > 0}$) will in general induce concentration jumps at the interface.

The currents of substrates and products\footnote{In analogy to the enzyme currents [Eq.~\eqref{eq:effective_euler-a}], we added integration constants to ensure that the substrate and product currents vanish far away from the droplet.} 
in Eq.~\eqref{eq:system_comoving},
\begin{subequations}
\label{eq:reactant_interface_currents}
\begin{align} 
    \mathbf{J}_s &= 
    - D \, \boldsymbol\nabla s - \Lambda \, s \, \chi_s \boldsymbol\nabla c - \mathbf{v} \, (s-s_\infty) 
    \,, \\
    \mathbf{J}_p &=  
    - D \, \boldsymbol\nabla p - \Lambda \, p\, \chi_p \boldsymbol \nabla c -\mathbf{v} \, (p-p_\infty) 
    \,,
\end{align}
\end{subequations}
must not only be continuous to conserve mass but also be finite-valued at the droplet interface.
This is not automatically guaranteed, because the concentration gradient $\boldsymbol\nabla c$ is singular at the interface.
To enforce boundedness for the currents and mesoscopic velocities, we now integrate $\mathbf{J_s} / s$ and $\mathbf{J_p}/p$ [cf.~Eq.~\eqref{eq:reactant_interface_currents}] over an infinitesimal line segment which crosses the sharp droplet interface.
After calculating this integral and taking the length of the line segment to zero, all except the first two terms on the right-hand side of Eq.~\eqref{eq:reactant_interface_currents} vanish. 
This leads to the boundary conditions
\begin{subequations}
\label{eq:concentration_jump}
\begin{align}
    \label{eq:concentration_jump_substrate}
    \frac{s\vert_\text{in}}{s\vert_\text{out}} 
    &= 
    \exp \biggl[ -\frac{\Lambda \, \chi_s \, \Delta c}{D} \biggr] \, , 
    \\
    \label{eq:concentration_jump_product}
    \frac{p\vert_\text{in}}{p\vert_\text{out}} 
    &= 
    \exp \biggl[ -\frac{\Lambda \, \chi_p \, \Delta c}{D} \biggr] 
    \, .
\end{align}
\end{subequations}
Thus, for finite $\Lambda$, the substrate and product concentration profiles in general exhibit a jump at the droplet interface.

When the interactions between enzymes, substrates, and products are sufficiently weak (${\lvert \Lambda \, \chi_{s,p} \, \Delta c / D \rvert \ll 1}$), the system reverts to the nonreciprocal limit discussed in our previous work~\cite{Demarchi2023}.
In this limit, the concentration profiles of substrates and products at the droplet interface become continuous.
In contrast, strong attractive enzyme-substrate and enzyme-product interactions ($\chi_{s,p} \ll 0$) will lead to an exponential enrichment of both substrates and products inside of the condensate.
We expect droplet motion to cease once the redistribution of substrates and products driven by these reciprocal interactions outweighs the redistribution caused by reactions and diffusion, i.e., when the system is close enough to thermodynamic equilibrium.

Taken together, conservation of mass dictates that the net flows of substrates and products at the droplet interface are continuous and finite, and 
amounts to Robin boundary conditions specified in  Eqs.~\eqref{eq:concentration_kink}~and~\eqref{eq:concentration_jump}. 
This allows us to revisit our previous analysis~\cite{Demarchi2023} for the more general case of ${\Lambda \geq 0}$.

To that end, one could directly solve Eq.~\eqref{eq:system_comoving_subdomain} in each subdomain (inside and outside the condensate), connect these solutions at the droplet interface by using Eqs.~\eqref{eq:concentration_kink}~and~\eqref{eq:concentration_jump}, and impose no-flux conditions in the far field.
However, one must also ensure that the solution conserves the total mass $n_\infty$ of substrates and products:
\begin{equation}
\label{eq:mean_concentration_condition}
    n_\infty 
    = 
    \langle s + p \rangle 
    =
    \frac{1}{|\Omega|} \int_{\Omega}\! d^dz \, [s(\mathbf{z})+p(\mathbf{z})]  \, ,
\end{equation}
where $|\Omega|$ is the total volume of the integration domain. 
For an open domain, where $|\Omega|\to\infty$, this implies that 
\begin{equation}
\label{eq:far_field_concentration_condition}
    \lim_{|\mathbf{z}|\to\pm\infty} [s(\mathbf{z})+p(\mathbf{z})] = n_\infty \, ,
\end{equation}
far away from the droplet.
At first glance, this additional constraint may seem trivial because the total concentration of substrates and products, ${n(\mathbf{x}) \coloneqq s(\mathbf{x}) + p(\mathbf{x})}$, is spatially uniform\footnote{Note that for ${\Lambda = 0}$, the total concentration follows a diffusion equation.} for ${\Lambda = 0}$.
However, this is in general not the case for ${\Lambda > 0}$ and thus requires separate consideration.
Specifically, as we will show next, a moving droplet can spatially redistribute reactant mass ${n(\mathbf{x},t)}$ in the laboratory frame.
Such mass redistribution is a key feature and control process of biochemical pattern-forming systems~\cite{Review::Halatek2018, Halatek2018, Brauns2020}.

\subsection{Redistribution of total reactant mass by moving droplets}
\label{sec:reciprocal_redistribution_reactants}

To analyze the profile of the total concentration of substrates and products, we add Eqs.~\eqref{eq:substrates_comoving_subdomain}~and~\eqref{eq:products_comoving_subdomain}, which gives
\begin{equation}
\label{eq:advection_diffusion_total_mass}
    0 
    = 
    \boldsymbol\nabla \cdot 
    \left( n \, \mathbf{v}  + D \, \boldsymbol\nabla n   
    \right) \,.
\end{equation}
Because of mass conservation, the reaction terms have cancelled out implying that, in the co-moving frame, the sum of the advective and diffusive mass fluxes of the total amount of reactants is divergence-free. 
Moreover, far away from the droplet, the total reactant concentration approaches the far-field value, ${\lim_{|\mathbf{z}|\to\infty} n(\mathbf{z}) = n_\infty}$, and becomes homogeneous, ${\lim_{|\mathbf{z}|\to\infty} \boldsymbol\nabla n(\mathbf{z}) = 0}$.
Finally, note that Eq.~\eqref{eq:advection_diffusion_total_mass} is valid both inside and outside of the condensate, but not at the droplet interface where $c(\mathbf{z})$ is singular.
Thus, each of these domains need to be analyzed separately and then connected by the Robin boundary conditions derived previously.
If the substrate and product concentration fields are smooth and continuous (${\Lambda = 0}$), then Eq.~\eqref{eq:advection_diffusion_total_mass} has only ${n(\mathbf{z}) = n_\infty}$ as solution.
In general, however, the substrate and product concentration fields are neither smooth nor continuous (${\Lambda > 0}$).
Since this complicates the analysis considerably, we focus on 1D systems in the following.

In a 1D geometry, the two droplet interfaces define the boundaries between three spatial subdomains that we label with indices ${i\in\{-,0,+\}}$, which represent the solvent to the left of the droplet ($-$), the droplet ($0$), and the solvent to the right of the droplet ($+$). 
Solving Eq.~\eqref{eq:advection_diffusion_total_mass}, gives for the total concentration of substrates and products in each subdomain
\begin{equation}
\label{eq:total_concentration_general}
    n(z) = C_i + \Delta C_i \, \exp(-v z / D) \, ,
\end{equation}
where $C_i$ and $\Delta C_i$ are integration constants.
To determine these constants, we will now and in the following, without loss of generality, assume that ${v \geq 0}$.
Because the concentration profiles must remain finite in the far field ${z \to \pm \infty}$, one has ${\Delta C_- = 0}$ for the solvent domain left to the droplet.
Furthermore, the far-field conditions [Eq.~\eqref{eq:far_field_concentration_condition}] imply that ${C_{\pm} = n_\infty}$.
By summing Eqs.~\eqref{eq:concentration_kink_substrate}~and~\eqref{eq:concentration_kink_product}, one finds Robin boundary conditions for the total reactant concentration at the two droplet interfaces at ${z = \pm R}$, which lead to the conclusion: ${C_0 = C_\pm = n_\infty}$.
Taken together, the total reactant concentration is given by
\begin{equation}
\label{eq:total_concentration_specified}
    n(z) = n_\infty + 
    \exp(- v z/D) \color{gray}\times\color{black}
    \begin{cases}
        0 \,,  & z < -R \, , \\
        \Delta C_0 \,, & |z| \leq R \, , \\
        \Delta C_+ \,, & z > R \, .
    \end{cases}
\end{equation}
The remaining integration constants, $\Delta C_0$ and $\Delta C_+$, cannot be determined without taking into account the concentration jumps of the substrates and products, Eq.~\eqref{eq:concentration_jump}.
Thus, they require resolving the corresponding concentration profiles, as will be discussed in the following.

\subsection{Inhomogeneous Helmholtz equation determines concentration profiles}

Using mass conservation, ${n(z) = s(z) + p(z)}$, to eliminate 
the concentration of products from Eq.~\eqref{eq:system_comoving_subdomain}, we arrive at an inhomogeneous Helmholtz equation with advection which describes the concentration profile of substrates in each subdomain,
\begin{equation}
    \label{eq:substrates_comoving_reduced}
    0 
    = 
    \partial_z  
    \left( v \, s + D \, \partial_z s   
    \right) - (k_1 \, c_\pm + k_2) \, s + k_2 \, n \,.
\end{equation}
Recall that $c_\pm$ refer to the enzyme concentrations inside and outside the droplet, respectively.
Also note that the total reactant concentration $n(z)$ is spatially inhomogeneous, as specified by Eq.~\eqref{eq:total_concentration_specified}.
The boundary conditions for the substrate profiles at the droplet interfaces are given by Eq.~\eqref{eq:concentration_kink_substrate} and Eq.~\eqref{eq:concentration_jump_substrate}.
These allow us to determine the distribution of substrates $s(z)$ and the total reactant concentration profile $n(z)$ up to two constants $\Delta C_0$ and $\Delta C_+$.
To now constrain these two constants, we turn to the distribution of products, $p(z) = n(z) - s(z)$.
In particular, the product concentration jumps at the two droplet interfaces, defined by Eq.~\eqref{eq:concentration_jump_product}, specify the remaining two constants $\Delta C_0$ and $\Delta C_+$.
By solving\footnote{Specifically, we used the computer algebra system Mathematica~\cite{Mathematica} to solve Eq.~\eqref{eq:substrates_comoving_reduced} within each subdomain; the full expressions can be found in Ref.~\cite{CodeGithub}.} Eq.~\eqref{eq:substrates_comoving_reduced}, we determined the
concentration profiles for substrates and products when the droplet was either moving (${v > 0}$) or stationary (${v = 0}$).

Our theoretical results for the concentration profiles align closely with our simulations, as shown in Fig.~\ref{fig:stationary_profiles_reciprocal} for a stationary droplet consisting of enzymes. 
As we discussed in Sec.~\ref{sec:model_reactions}, near the center of the droplet, enzyme-catalyzed reactions increase the concentration of products at the expense of substrates.
These reaction-induced concentration gradients correspond to particle fluxes which bring substrates towards the condensate and transport products away.
In addition, attractive enzyme-substrate and enzyme-product interactions increase the concentration of both substrates and products by a discontinunous concentration jump at the droplet interface [Eq.~\eqref{eq:concentration_jump}].
In Secs.~\ref{sec:reciprocal_propulsion_instability}--\ref{sec:analysis_role_reciprocity}, we explore the implications of these concentration profiles on droplet self-propulsion. 

\section{Evaluating the concentration differences of substrates and products}
\label{appendix:ambiguity_concentration_difference}

Note that there is a subtlety when evaluating the concentration differences of the substrates $\Delta s(v)$ and the products $\Delta p(v)$ between the two droplet interfaces. 
Since we are in the sharp interface limit, it is not entirely clear when evaluating the integrals 
\begin{subequations}
\label{eq:integral_ambiguity}
\begin{align}
    \Delta s(v) &\coloneqq \int_{\mathcal{D}} \! d z \, \partial_z s(z) 
    \, , \\
    \Delta p(v) &\coloneqq \int_{\mathcal{D}} \! d z \, \partial_z p(z) 
    \, ,
\end{align}
\end{subequations}
which value we should choose for the lower and upper limits.
These could be the values at the inner or outer boundaries or an interpolation between the two.
The difference between these choices arises from the discontinuities in the concentration profiles which, taking the substrate concentrations at the right droplet interface as an example, imply ${\lim_{\varepsilon\to 0} s(R-\varepsilon) \neq \lim_{\varepsilon\to 0} s(R+\varepsilon)}$.
Currently, we cannot resolve which is the correct choice solely based on theoretical arguments.
Hence, we proceeded heuristically and compared the various choices with simulation data for the droplet speed as a function of the enzyme mobility and the reaction rates.
It turned out that these simulation data are only compatible with choosing a droplet domain that excludes its boundary, ${\mathcal{D} = \{z : |z| < R\}}$, which for the enzyme concentration profile implies ${c(z) = c_+ \, \forall \, |z| < R}$ else $c_-$.
In this case, one takes the concentration values at the inner sides of the droplet interfaces when evaluating Eq.~\eqref{eq:integral_ambiguity}.
This ambiguity could be resolved in future studies by using an enzyme concentration profile with a smooth interface, or by using a kinked (i.e., with a discontinuity in the gradient) enzyme concentration profile~\cite{JinFisher_PRB1993}. 
Notwithstanding these limitations of the sharp interface approximation, as we show in Fig.~\ref{fig:velocities_lambda}, it quantitatively predicts the droplet velocity in agreement with our simulation results.

\section{Reciprocity suppresses droplet self-propulsion for attractive enzyme-substrate and enzyme-product interactions}
\label{appendix:large_lambda}
In the following, we repeat the calculation performed in Sec.~\ref{sec:strong_reciprocity}, but now for the case ${\chi_s < \chi_p < 0}$.
The concentration jumps for substrates and products at the droplet interface, Eq.~\eqref{eq:concentration_jump}, in the limit ${\Lambda \, \epsilon_0 / D \rightarrow \infty}$, are then determined by:
\begin{subequations}
\begin{align}
    \frac{s\vert_\text{in}}{s\vert_\text{out}} 
    &= 
    \exp \biggl[ -\frac{\Lambda \, \chi_s \, \Delta c}{D} \biggr] \equiv \Omega_s
    \to \infty
    \, , 
    \\
    \frac{p\vert_\text{in}}{p\vert_\text{out}} 
    &= 
    \exp \biggl[ -\frac{\Lambda \, \chi_p \, \Delta c}{D} \biggr] \equiv \Omega_p
    \to \infty
    \, .
\end{align}
\end{subequations}
In this case, the slope of the maximal chemical potential imbalance at ${v=0}$ has the modified [cf. Eq.~\eqref{eq:maximal_chemical_potential_imbalance_strong_reciprocity}] asymptotic form
\begin{equation}
    \partial_v \Delta\mu_0(v) \rvert_{v=0} = \Omega_p \frac{2 R n_\infty}{D} \frac{l_+^2 (\chi_p-\chi_s)-l_0^2 \chi_p}{l_+^2-l_0^2} \, .
\end{equation}
Since ${\chi_s < \chi_p < 0}$ and ${l_\pm < l_0}$, it follows that ${\partial_v \Delta\mu_0(v)\rvert_{v=0} \to -\infty}$, and thus that the state ${v=0}$ is stable. 
The system could still be in the bi-stable regime, but further analytic calculations are unfeasible.

\section{Positioning and coexistence}
\label{appendix:positioning_coexistence}

As we have shown in the present work, the presence of reciprocal interactions (${\Lambda > 0}$) further constrains the parameter regime in which droplet self-propulsion can be observed.
This raises the question about condensate positioning, coexistence, and divisions, which we have reported previously~\cite{Demarchi2023}.
As shown in Supplemental Video~5, condensates position themselves to the center of their confinement, which leads to a system configuration with maximal symmetry.
Similarly, droplets can also show coexistence or even elongate and divide [Supplemental Video~6 and Supplemental Video~7].
Thus, the presence of reciprocal interactions does not qualitatively change these dynamics.

\section{Supplemental Videos}

\textbf{Supplemental Video 1 (\path{video_1_condensate_fluxes_no_force.mp4}).}
Chemical potential profile and enzyme currents under the condition that the 3D condensate moves with a defined speed, which is varied in the movie. 
In this video there is no non-equilibrium driving force. 
Stable fixed point is indicated by a filled circle.
Parameters:
${c_-=0.5 c_+}$, ${R=l_0}$, ${k_1=0 k_2 / c_+}$, ${\Lambda=0}$, and ${s+p=c_+}$, ${\Delta\chi\coloneqq \chi_p - \chi_s = 0 r}$ and ${M(c) = M c}$ with ${M=10 D / \epsilon_0}$.

\textbf{Supplemental Video 2 (\path{video_2_condensate_fluxes_with_reactions.mp4}).}
Chemical potential profile and enzyme currents under the condition that the 3D condensate moves with a defined speed, which is varied in the movie. 
In this video, the condensate experiences a driving force due to its interactions with the non-uniformly distributed substrates and products. 
Stable fixed points are indicated by filled circles.
Parameters:
${c_-=0.5 c_+}$, ${R=l_0}$, ${k_1=2 k_2 / c_+}$, ${\Lambda=0}$, ${s+p=c_+}$, ${\Delta\chi\coloneqq \chi_p - \chi_s = 4 r}$ and ${M(c) = M c}$ with ${M=10 D / \epsilon_0}$.

\textbf{Supplemental Video 3 (\path{video_3_condensate_dissolve.mp4}).}
Example where a self-propelling 2D condensate is only metastable.
After an initial transient period of self-propulsion, the condensate dissolves.
Compared to the initial conditions discussed in Appendix~\ref{appendix:initial_conditions}, the low-concentration phase was further diluted by $9.1\%$.
Parameters: ${c_-=0.1 c_+}$, ${k_1=k_2 / c_+}$, ${\chi_s = -0.05 r}$, ${\chi_p = -0.01 r}$, ${\Lambda=0}$, ${w = 0.1 l_0}$, ${R = l_0}$, ${s+p = c_+}$, and ${M(c) = M c}$ with ${M=1000 D / \epsilon_0}$.
The circular domain has radius ${L = 7 l_0}$.

\textbf{Supplemental Video 4 (\path{video_4_condensate_propel.mp4}).}
Example where a self-propelling 2D condensate is stable.
The condensate does not dissolve.
Compared to the initial conditions discussed in Appendix~\ref{appendix:initial_conditions}, the low-concentration phase was not diluted any further.
Parameters: ${c_-=0.1 c_+}$, ${k_1=k_2 / c_+}$, ${\chi_s = -0.05 r}$, ${\chi_p = -0.01 r}$, ${\Lambda=0}$, ${w = 0.1 l_0}$, ${R = l_0}$, ${s+p = c_+}$, and ${M(c) = M c}$ with ${M=1000 D / \epsilon_0}$.
The circular domain has radius ${L = 7 l_0}$.

\textbf{Supplemental Video 5 (\path{video_5_reciprocal_interactions.mp4}).}
Examples of 1D condensates which self-propel, position themselves in a container, and control their size even in the presence of reciprocal interactions.
Parameters: ${c_-=0.1 c_+}$, ${k_1=k_2 / c_+}$, ${\chi_s = -0.05 r}$, ${\chi_p = -0.01 r}$, ${w = 0.1 l_0}$, ${R = l_0}$, ${s+p = c_+}$, and ${M(c) = M c}$.
For the simulation showing self-propulsion: ${M=5000 D / \epsilon_0}$ and ${\Lambda=4 D / \epsilon_0}$.
For the simulation showing positioning and coexistence: ${M=100 D / \epsilon_0}$ and ${\Lambda=20 D / \epsilon_0}$.
The linear domain has size ${L = 30 l_0}$ for the self-propulsion example, $L = 3 l_0$ for the positioning example, or $L = 5 l_0$ for the coexistence example.

\textbf{Supplemental Video 6 (\path{video_6_reciprocal_division_lmbd_0.1}).}
Example of a condensate, for weak reciprocity parameter ${\Lambda=0.1 D / \epsilon_0}$, which divides upon switching the catalysis rate from $k_1 = 100 \, k_2/c_+$, for which the droplet is stable, to $k_1 = 1 \, k_2/c_+$.
The two smaller droplets remain stable when the catalysis rate is increased again.
Simulations were performed in a 3D cylindrical domain (radius $L_r = 1.5 l_0$ and half-height $L_z = 4 l_0$).
Parameters: $M = 10 D/\epsilon_0$, $\chi_s = -0.5 r$, $w = 0.05 l_0$.

\textbf{Supplemental Video 7 (\path{video_6_reciprocal_division_lmbd_1}).}
Example of a condensate, for reciprocity parameter ${\Lambda=1.0 D / \epsilon_0}$, which elongates upon switching the catalysis rate from $k_1 = 100 \, k_2/c_+$, for which the droplet is stable, to $k_1 = 1 \, k_2/c_+$.
The elongated droplet then divides into two smaller droplets once the catalysis rate is increased again.
Simulations were performed in a 3D cylindrical domain (radius $L_r = 1.5 l_0$ and half-height $L_z = 4 l_0$).
Parameters: $M = 10 D/\epsilon_0$, $\chi_s = -0.5 r$, $w = 0.05 l_0$.

\end{document}